\documentclass[%
preprint,
superscriptaddress,
showpacs,
amsmath,
amssymb,
onecolumn,
longbibliography,
prl
]{revtex4-2}

\usepackage{amsmath}
\usepackage{amsfonts}
\usepackage{amssymb}
\usepackage{array}
\usepackage{bbm}
\usepackage{bm}
\usepackage{bbold}
\usepackage{braket}
\usepackage{bigints}
\usepackage{dsfont}
\usepackage[dvipsnames]{xcolor}
\usepackage{epsfig}
\usepackage{epstopdf}
\usepackage{graphicx}
\usepackage{multirow}
\usepackage{mathtools}
\usepackage[T1]{fontenc}
\usepackage{times} 
\usepackage{tensor}
\usepackage[utf8]{inputenc}
\usepackage[unicode=true,breaklinks=true,colorlinks=true]{hyperref}

\usepackage[normalem]{ulem}
\usepackage{pdfpages}
\usepackage{pgffor}

\hypersetup{citecolor=blue,urlcolor=blue}

\makeatletter 
\AtBeginDocument{\let\LS@rot\@undefined} 
\makeatother

\newcommand{\rev}[1]{\textcolor{black}{#1}}
\newcommand{\lxb}[1]{\textcolor{black}{#1}}

\begin{document}

\title{Quantum metrological capability as a probe for quantum phase transition}

\author{Xiangbei Li}
\affiliation{School of Physics, Hubei Key Laboratory of Gravitation and Quantum Physics, International Joint Laboratory on Quantum Sensing and Quantum Metrology, Institute for Quantum Science and Engineering, Huazhong University of Science and Technology, Wuhan 430074, China}
\author{Yaoming Chu}
\email{yaomingchu@hust.edu.cn}
\author{Shaoliang Zhang}
\author{Jianming Cai}
\email{jianmingcai@hust.edu.cn}
\affiliation{School of Physics, Hubei Key Laboratory of Gravitation and Quantum Physics, International Joint Laboratory on Quantum Sensing and Quantum Metrology, Institute for Quantum Science and Engineering, Huazhong University of Science and Technology, Wuhan 430074, China}

\begin{abstract}
The comprehension of quantum phase transitions (QPTs) is considered as a critical foothold in the field of many-body physics. Developing protocols to effectively identify and understand QPTs thus represents a key but challenging task for present quantum simulation experiments. Here, we establish a dynamical quench-interferometric framework to probe a zero-temperature QPT, which utilizes the evolved state by quenching the QPT Hamiltonian as input of a unitary interferometer. The metrological capability quantified by the quantum Fisher information shows a unique peak in the vicinity of the quantum critical point, allowing us to probe the QPT without cooling the system to its ground state. We show that the probing can be implemented by extracting quantum fluctuations of the interferometric generator as well as parameter estimation uncertainty of the interferometric phase, and subsequently allows identifying the boundary of the phase diagram. Our results establish an important link between QPTs and quantum metrology, and enrich the toolbox of studying non-equilibrium many-body physics in current quantum simulators.
\end{abstract}

\maketitle

\date{\today}

\maketitle

\newpage

\noindent
\textbf{Introduction}\\
\noindent
Quantum phase transitions (QPTs) represent a vigorous subject strongly relevant in various fields, ranging from condensed matter to quantum field theory and cosmology~\cite{sachdev2011quantum, MatthiasVojta_2003}. It is commonly considered an essential ingredient for our perception of a plethora of fascinating quantum phenomena in strongly correlated many-body physics such as quantum magnetization~\cite{sachdev2008quantum}, spin liquid~\cite{Savary_2017}, topological order~\cite{Wen2013_topological_order} and superconductor-insulator transition~\cite{Gantmakher_2010}. In the last few years, the booming development of AMO physics and quantum simulation technologies opens a pristine and versatile experimental avenue towards the study of QPTs through controllable realizations of a variety of theoretical models \cite{zhang2017observation, Georgescu2014_RMP, Dingshun2018prx, Harris2018science,Daley2022quantumsimulation, Duan2024trappedions, Pan2024_antiferromagnetic}. This also naturally inspires intensive attention of synergistically investigating QPTs in quantum simulators beyond the traditional Landau-Ginzburg paradigm, by resorting to the concepts and strategies developed in quantum information science~\cite{osterloh2002scaling, Gu2010, CAROLLO20201, 2024Arnold_prl, arnold2023machinelearningphasetransitions}.

Remarkably, a quantum-fidelity approach to QPTs has been put forward to unveil sharp variations of the ground state close to critical point from a universal information-theoretic picture~\cite{Zanardi2006_PRE, ZanardiPaolo2007, Zanardi2007_prl, Gu2010, CAROLLO20201}. 
However, the requirement of cooling a complex many-body system to nearly zero temperature might be challenging. More recently, frameworks based on non-equilibrium dynamical response after sudden quantum quenches interestingly demonstrate the possibility of establishing a versatile toolbox for locating and witnessing QPTs \cite{2019PRL_Gong, 2021Das_PRX,2018Dora_PRL,2019Duan_prl,2023PRB_Halimeh, 2023PRB_Dutta, Aditi2024_PRA, paul2022hidden,  Nie2020_prl, 2021Sun_PRB, 2018Silva_prl, Jafari2020_PRA, kobayashi2024quantum}, and are expected to become a routine practice. While being compelling, efficient extraction of the quantities that act as indicators of the QPTs in those previous frameworks, such as short-range correlations \cite{2019PRL_Gong,2021Das_PRX}, out-of-time-order correlators \cite{2018Dora_PRL,2019Duan_prl}, entanglement entropy \cite{2021Das_PRX} and even multi-site entanglement \cite{Aditi2024_PRA}, typically requires the ability of implementing local measurements or other complex operations (e.g.\,time-reversed evolution) on a generic many-body system. Therefore, further strengthening and enriching the toolbox of probing QPTs with quantum information strategies  still remains largely unexplored and deserves intense efforts.

In this Letter, we establish an efficient framework for probing QPTs by integrating nonequilibrium quench dynamics with the concept of interferometry in quantum metrology. Specifically, we consider the post-quenched state prepared by time evolution governed by a QPT Hamiltonian as input of a unitary interferometer. \lxb{We stress that our interferometric framework is different from the typical scenarios of critical metrology, which focus on estimating the QPT’s control parameter that drives the phase transition \cite{Zanardi2006_PRE,2008Zanardi_pra, Invernizzi2008PRA, Mehboudi2016PRA, 2018prx_Marek, 2021Chu_prl}.} The {\it metrological capability} of the state, theoretically quantified by the quantum Fisher information, determines the ultimate precision of estimating the interferometric phase. As our main result, we find that the metrological capability interestingly shows a peak around the quantum critical point (QCP). We further demonstrate that this enhanced metrological capability can be effectively accessed through quantum fluctuation of the interferometric generator or the estimation uncertainty of the phase, which merely requires the measurement of certain collective spin operators by global detection of the QPT system.
As a direct application, we obtain the boundary of the QPT phase diagram and show that it coincides well with the theoretical prediction of equilibrium QCPs. Our framework not only provides a new promising and experimentally practical avenue to investigate QPTs and the comprehension of non-equilibrium many-body quantum phenomena, but also reveals further deep connections between quantum metrology and quantum criticality \cite{2008Zanardi_pra,Invernizzi2008PRA, Mehboudi2016PRA,2018prx_Marek, You2017_Science, 2018Tommaso_prl, Chu2020_prl, 2021Chu_prl, ding2022enhanced, 2024Hotter_prl, Braun2018_RMP, Lee2020, 2023Chu_prl, 2024Yu_nsr, Guan2024_prl,Block2024Scalable, Mihailescu_2024Multiparameter, mihailescu2024uncertainquantumcriticalmetrology, adani2024criticalmetrologyminimallyaccessible}.

\vspace*{2ex}
\noindent
\textbf{Quench-interferometric framework for probing QPTs}\\
\noindent
At the heart of quantum metrology lies the quantum Fisher information (QFI) ~\cite{Caves1994_prl,2018RMP_Pezze, Liu_2020}, which \rev{can} quantify the metrological capability of a probe state with respect to the unitary parameterization generated by a Hermitian operator $\hat{\mathcal{O}}$. \lxb{In this work, we focus on the standard interferometric framework \cite{2006prl_quantum_metrology,giovannetti2011advances,Wineland1994PRA, 2018Tommaso_prl,Zoller2016measuring}, where  a local $N$-particle generator $\hat{\mathcal{O}}=\sum_{i=1}^N \hat{O}_i$ is considered}. Given a pure probe state $\ket{\Psi}$, the associated QFI can be easily computed or measured from quantum fluctuation of the generator \lxb{\cite{Zoller2016measuring,Liu_2020}}
\begin{equation}
F_Q[\ket{\Psi},\hat{\mathcal{O}}]=4\left[\langle \Psi|\hat{\mathcal{O}}^2|\Psi\rangle-\langle \Psi|\hat{\mathcal{O}}|\Psi\rangle^2\right].
\label{eq:FQ_Var}
\end{equation}
This form of QFI is elegantly linked to the connected correlation function, resulting in significant implications well beyond the scope of quantum metrology. Recently, in paradigmatic spin models of QPTs, it has been demonstrated that the QFI of the ground state can exhibit superextensive divergence or abrupt changes in the vicinity of a QCP \cite{Zoller2016measuring,2018Tommaso_prl,Niezgoda2021}. Such fundamentally interesting behaviors could unveil a variety of  fascinating phenomena such as the witness of critical multipartite entanglement \cite{Zoller2016measuring}, the redistribution of quantum noise among extensive observables \cite{2018Tommaso_prl} and the emergence of many-body nonlocality \cite{Niezgoda2021}, which however is hindered by the requirement of ground state preparation.

%
\begin{figure}[t] 
\centering 
\includegraphics[width=8.6cm]{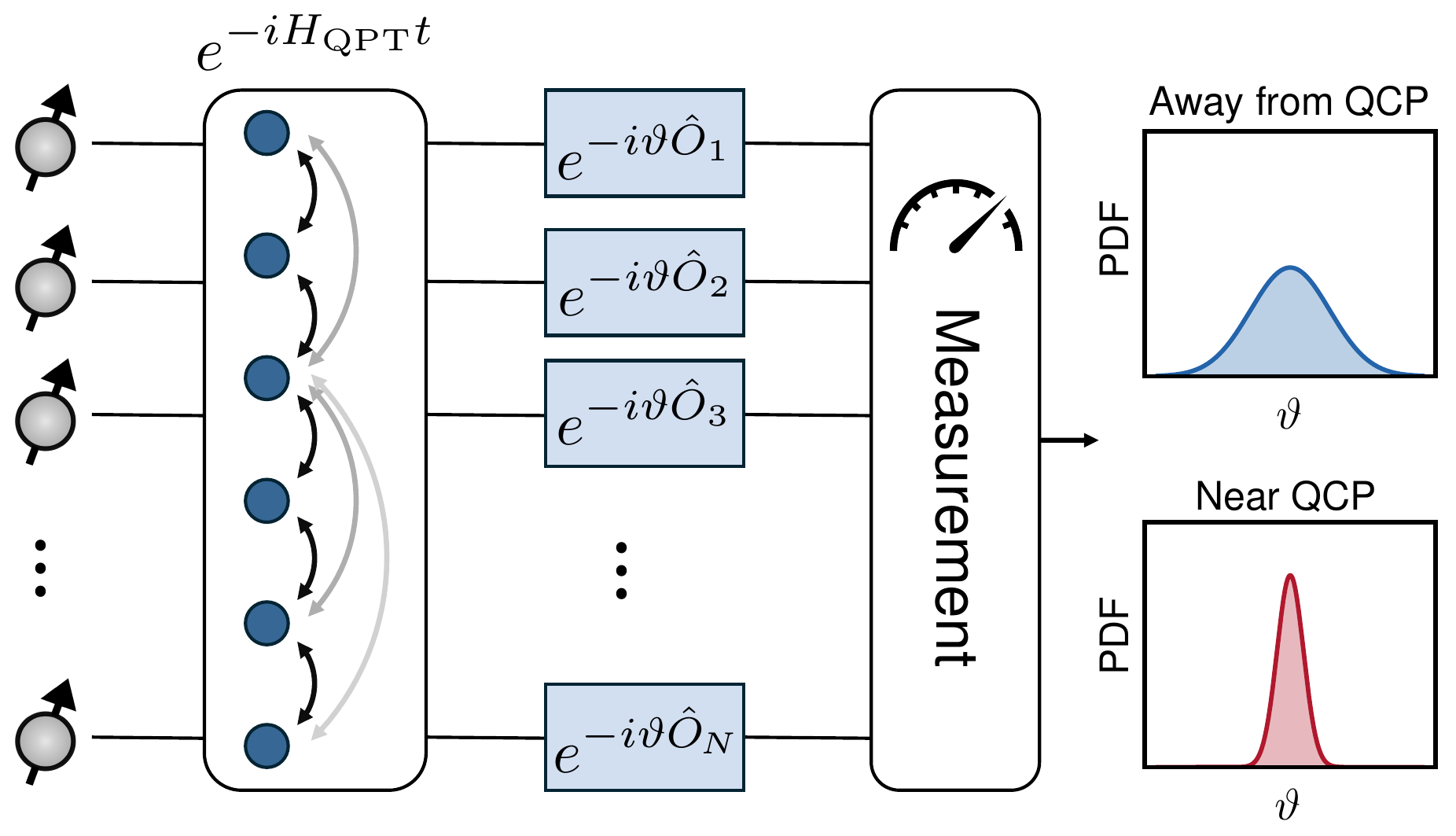} 
\caption{\textbf{Schematic graph of the quench-interferometric framework for probing QPTs}. A many-body system of $N$ particles is prepared into a probe state via nonequilibrium quantum dynamics, achieved by suddenly quenching the QPT Hamiltonian $H_{\mathrm{QPT}}$. This probe state is then used to estimate an unknown phase $\vartheta$ encoded by a local generator $\hat{\mathcal{O}}=\sum_{i=1}^N \hat{O}_i$. The metrological capability of the probe state can be significantly enhanced when the quenched Hamiltonian approaches the quantum critical point, as evidenced by a much narrower probability density function (PDF) of the estimated value of $\vartheta$.}
\label{fig1} 
\end{figure}

Here, we consider the QFI of a probe state prepared through nonequilibrium quantum dynamics by suddenly quenching the QPT Hamiltonian, thus circumventing the challenging preparation of a many-body system near its ground state. It is worth to remark that the time-dependent QFI evolution in quench dynamics might open new possibilities beyond the knowledge of quantum systems at equilibrium, enabling alternative pathways towards various intriguing phenomena such as entangling dynamics \cite{Chu2024,2008Amico_RMP_Entanglement, 2017Adam_EntanglementGrowth,2023Chu_prl, Guan2024_prl,Baykusheva2023}, transient quantum coherence \cite{Garttner2018,LewisSwan2020} and the interplay between equilibrium and dynamical QCPs \cite{2019PRL_Gong}. Explicitly, our study
proposes a quench-interferometric framework that employs post-quenched states as input of interferometers, namely
\begin{equation}
\label{Eq:framework}
|\Psi_t(\vartheta)\rangle=e^{-i\vartheta \hat{\mathcal{O}}} e^{-iH_{\mathrm{QPT}}t}|\Psi_0\rangle.
\end{equation}
We firstly evolve the system to a probe state by a QPT Hamiltonian, i.e. $|\Psi_t\rangle=e^{-iH_{\mathrm{QPT}}t}|\Psi_0\rangle$, and then input it into a quantum interferometer with a phase unitarily encoded by $e^{-i\vartheta \hat{\mathcal{O}}}$, see Fig.\,\ref{fig1}.
The ultimate measurement precision of the interferometric phase $\vartheta$ is determined by the QFI in $|\Psi_t(\vartheta)\rangle$, which is uniformly (namely, independent of $\vartheta$) given by $F_Q(t)=F_Q[|\Psi_t\rangle,\hat{\mathcal{O}}]$.

\begin{figure*}[t] 
\centering 
\includegraphics[width=16cm]{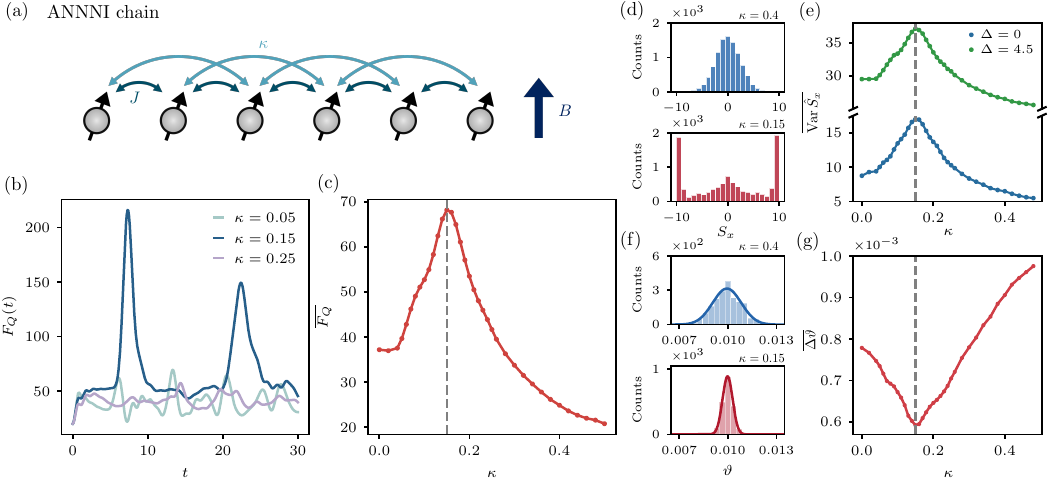} 
\caption{\textbf{Probing the zero-temperature QPT of the ANNNI model}. (a) The ANNNI chain consists of nearest-neighboring and next-nearest-neighboring Ising interaction as well as a transverse biased field, denoted by the parameters $(J,\kappa, B)$ respectively. In our simulation we set $J=1$ and $B=0.75$, which leads to a QCP at $\kappa_c\approx 0.15$ marked by the vertical dashed gray lines in the following panels.  (b) The QFI dynamics for different values of $\kappa$. The apparently enhanced QFI (dark blue curve) corresponds to the quench parameter $\kappa$ close to the QCP.
(c) The time-averaged QFI exhibits a maximum in the vicinity of the QCP. 
(d) Histogram of $m=10000$ simulated measurement outcomes of $\hat{S}_x$ on the posted-quenched state at $t=7.5$. 
(e) The time average of quantum fluctuations in Eq.\,\eqref{Eq: sample variance} is obtained for different measurement resolutions. They all show a peak that coincides well with the equilibrium QCP. 
(f) Histogram of 2000 simulated  MLE of the interferometric phase using the posted-quenched state at $t=7.5$. The solid lines represent the Gaussian fitting centered around the true value $\vartheta=0.01$.
(g) Time-averaged uncertainty of MLE estimate, see Eq.\,\eqref{Eq:AveUncertainty}, identifies the QCP by its minimum.
Here, the system size is set as $N=20$ and the window of the time-averaging operation is chosen as $T=15$.}
\label{fig2} 
\end{figure*}

Below, we investigate identification of the QPT by analyzing QFI evolution of the post-quenched state in Eq.\,\eqref{Eq:framework}. As an illustrative example, we focus on the zero-temperature paramagnetic-to-ferromagnetic QPT embedded in the anisotropic next-nearest-neighbor Ising (ANNNI) chain, see Fig.\,\ref{fig2}(a), which is a representative model describing generic (quantum chaotic) many-body systems~\cite{2013Schuricht_prb, 2021Das_PRX}. The model Hamiltonian reads,
\begin{equation}
    H(\kappa,B) = -J\sum_{i=1}^{N} \sigma^x_i \sigma^x_{i+1} + \kappa \sum_{i=1}^{N} \sigma^x_i \sigma^x_{i+2} - B \sum_{i=1}^{N} \sigma^z_i,
\label{Eq: ANNNI Hamiltonian}
\end{equation}
where $\sigma^{x,z}_i$ are the Pauli operators at site $i$, and a periodic boundary condition is employed with $\sigma_{N+1}^x=\sigma_{1}^x$, $\sigma_{N+2}^x=\sigma_{2}^x$. We set $J=1$ in the following for simplicity. This model is generally nonintegrable \cite{2013Schuricht_prb,2017Alba_prl} except for the special scenario of $\kappa=0$. Second-order perturbation theory determines the paramagnetic-to-ferromagnetic transition line of the ANNNI chain as $1-2\kappa_c = B_c - \kappa_c B_c^2/(2-2\kappa_c)$~\cite{Beccaria2006_prb}. In our framework, we initialize this spin chain in a fully polarized state $|\Psi_0\rangle=|{\uparrow}\rangle^{\otimes N}$ and let it evolve under the ANNNI Hamiltonian, i.e. 
\begin{equation}
\label{Eq:Psit}
   \ket{\Psi_t}=e^{-i H(\kappa, B) t}\ket{\Psi_0}.
\end{equation}

We consider the collective spin operator $\hat{\mathcal{O}}=\hat{S}_x=\sum_{i=1}^N\sigma_i^x/2$ as the phase-translation generator. Numerical calculations of $F_Q(t)=F_Q[\ket{\Psi_t},\hat{S}_x]$ for 20 spins are performed using the exact diagonalization (ED) method \cite{quspin1,quspin2}. In Fig.\,\ref{fig2} (b-c), we demonstrate that the QFI evolution of $\ket{\Psi_t}$ prepared by non-equilibrium quench dynamics in Eq.\,\eqref{Eq:Psit}  can be employed to probe the QPT. It can be seen in Fig.\,\ref{fig2} (b) that the QFI dynamics close to the QCP is radically enhanced in contrast to the scenarios away from the QCP.  The averaged QFI is calculated over a time window of $T$ to quantify the overall difference across the critical point, i.e.
\begin{equation}
    \overline{F_Q}=(1/T)\int_0^T F_Q(t) \,\mathrm{d}t.
\end{equation}
Remarkably, $\overline{F_Q}$ displays a peak in the vicinity of the QCP as shown in Fig.\,\ref{fig2} (c), which serves an effective indicator to locate the QPT. \rev{We note that the emergence of such a peak can be related to the spontaneous symmetry breaking \cite{supplement}}. Our numeric analysis also shows that these results are robust to a finite system size and open boundary condition \cite{supplement}.

\vspace*{2ex}
\noindent
\textbf{Probing QPTs by global quantum fluctuations}\\
\noindent
\lxb{The quantum fluctuations of the generator, defined as $\operatorname{Var}[\hat{\mathcal{O}}]=\langle\Psi|\hat{\mathcal{O}}^2|\Psi\rangle-\langle \Psi|\hat{\mathcal{O}}|\Psi\rangle^2$, are proportional to the QFI [see Eq. \eqref{eq:FQ_Var}]. It thus reflects the metrological capability of the post-quenched state.}
Therefore, a direct probing of QPTs via the metrologically capability can be implemented by the measurement of $\hat{\mathcal{O}}$ (e.g. $\hat{S}_x$ in ANNNI example) with respect to the post-quenched state $\ket{\Psi_t}$. For a measurement repetitions of $m$ times, one obtains a series of random outcomes $\{ S_{x}^{(1)}, S_{x}^{(2)}, \ldots, S_{x}^{(m)}\}$ taking (half-)integer values in $[-N/2,N/2]$. The fluctuations of $\hat{S}_x$ can be efficiently estimated using the Bessel Formula,
\begin{equation}
\operatorname{Var}[\hat{S}_x] = \frac{1}{m-1} \sum_{i=1}^{m} \left( S_{x}^{(i)} - \langle \hat{S}_x \rangle \right)^2,
\label{Eq: sample variance}
\end{equation}
where $\langle \hat{S}_x\rangle=\sum_{i=1}^m S_{x}^{(i)}/m$. As demonstrated in Fig.\,\ref{fig2}(d), the measurement outcomes near the QCP apparently satisfy a wider probability distribution than the non-critical case. We calculate the time-averaged fluctuations $\overline{\operatorname{Var} \hat{S}_x}=(1/T)\int_0^T \operatorname{Var}[\hat{S}_x] \,\mathrm{d}t$, which evidently shows a peak versus the quench parameter $\kappa$ around the QCP, see Fig.\,\ref{fig2}(e). 

To account for the influence of a limited experimental measurement resolution that technically coarsens the measurement outcomes, the outcome $S_x^{(i)}$ for the $i$-th measurement can be assumed to be continuously distributed, with the corresponding measurement operator given by \cite{2016PRLDavis,2016PRL_Dur,2018RMP_Pezze, supplement}
\begin{equation}
\label{Eq:CM}
    \hat{\Pi}(S_x^{(i)}) = \frac{1}{\sqrt{2\pi} \Delta} e^{-(S_x^{(i)} - \hat{S}_x)^2 / (2\Delta^2)}.
\end{equation}
The standard deviation $\Delta$ effectively quantifies the measurement resolution, and when $\Delta\to0$, Eq.\,\eqref{Eq:CM} recovers the ideal projective measurement $\hat{\Pi}(S_x^{(i)}) =\delta(S_x^{(i)}-\hat{S}_x)$ with $\delta(x)$ denoting the Dirac delta function. We numerically simulate the measurement process of $\hat{S}_x$ under different resolutions and then evaluate quantum fluctuation based on Eq.\,\eqref{Eq: sample variance}. In Fig.\,\ref{fig2} (e), we depict the time-averaged $\overline{\operatorname{Var}\hat{S}_x}$ versus the quench parameter $\kappa$. Our results shows that the position of the peak at QCP is not influenced even for a poor measurement resolution of $\Delta\sim\sqrt{N}\approx4.5$, which would facilitate the observable $\operatorname{Var}[\hat{S}_x]$ to function as a trustworthy probe of QPTs.

\vspace*{2ex}
\noindent
\textbf{Probing QPTs by estimation uncertainty}\\
\noindent
Naturally, \rev{the quantum probe state with a larger metrological capability gives to  a better precision in estimating the phase $\vartheta$ associated with the unitary parameterization, i.e., $|\Psi_{t}(\vartheta)\rangle=e^{-i\vartheta \hat{\mathcal{O}}}|\Psi_t\rangle$.} 
Stated by the so-called quantum Cram{\'e}r-Rao bound, the best achievable estimation precision by detecting $n$ copies of $|\Psi_{t}(\vartheta)\rangle$ is lower bounded by $(\Delta \vartheta)^2 \geq 1/[n F_Q(|\Psi_t\rangle,\hat{\mathcal{O}})]$. 
It should be pointed out that implementing the optimal measurement strategy to saturate such a bound is usually challenging. Here, we consider the experimentally feasible observables and show that the corresponding non-optimal detection in estimating $\vartheta$ can also witness the QPT.

For a general measurement, the outcomes obeys a $\vartheta$-dependent probability distribution, which determines the maximal information that can be extracted about $\vartheta$, i.e. $(\Delta \vartheta)^2\geq 1/(n F_C)$, where $F_C$ denotes the Fisher information (FI) and $F_C\leq F_Q[|\Psi_t\rangle,\hat{\mathcal{O}}]$. This classical Cram{\'e}r-Rao bound can be achieved by the maximum likelihood estimation (MLE) for a sufficiently large $n$  \cite{fisher1925theory,Braunstein_1992,supplement}. In the explicit example of ANNNI model, we estimate $\vartheta$ by \rev{suitably choosing to measure} the global magnetization $\hat{S}_z = \sum_i \sigma_i^z/2$ \cite{supplement}. We numerically simulate the estimation uncertainty of $\vartheta$ by utilizing the MLE, which is constructed from 50000 measurement outcomes. In order to evaluate the estimation uncertainty $\Delta\vartheta$, namely the fluctuation of the unbiased MLE, we repeat each estimation by 2000 times, from which the distribution of the MLE is obtained in Fig.\,\ref{fig2} (f). It can be seen that the MLE near the QPT (corresponding to $\kappa\approx0.15$) shows a narrower distribution than non-critical cases, \rev{directly} indicating a better metrological performance achieved by the probe state generated at QCP. In Fig.\,\ref{fig2} (g), we further compute the time average of the estimation uncertainty, i.e.
\begin{equation}
\label{Eq:AveUncertainty}
    \overline{\Delta \vartheta} =\frac{1}{T} \int_0^{T} \Delta \vartheta \,\mathrm{d}t.
\end{equation}
It clearly exhibits a minimum around the QCP. Hence, the parameter estimation uncertainty achieved by the probe state prepared through non-equilibrium quench dynamics effectively identifies the QPT.
\begin{figure}[t] 
\centering 
\vspace{-0.2cm}
\includegraphics[width=8.8cm]{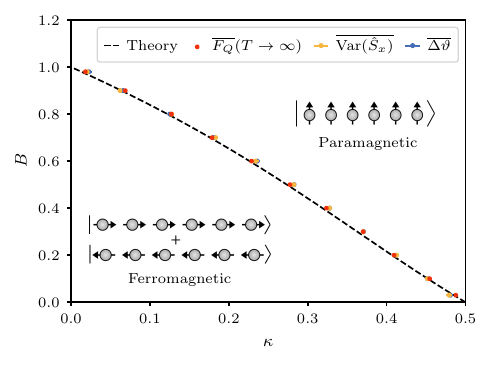} 
\caption{
\textbf{Identification of the phase diagram of the ANNNI model.} In the $(\kappa, B)$ parameter plane of the ANNNI chain, see Eq.\,\eqref{Eq: ANNNI Hamiltonian}, we numerically  determine the boundary of the phase diagram by finding the maximum of the time-averaged QFI and the global quantum fluctuation [$\Delta=4.5$ in Eq.\,\eqref{Eq:CM}], as well as the minimum of the time-averaged phase estimation uncertainty, see Fig.\ref{fig2}. The boundaries obtained based on our dynamical quench-interferometric framework all agree well with the theoretical prediction by the second-order perturbation theory.  Here, \lxb{we set $N=20$. The $\overline{\operatorname{Var}(\hat{S}_x)}$ and $\overline{\Delta\vartheta}$ are averaged over a finite time window $T=30$ with a sampling interval $\Delta t=0.3$ \cite{supplement}.}}
\label{fig3} 
\end{figure}

\vspace*{2ex}
\noindent
\textbf{The feasible detection of quantum phase diagram}\\
\noindent
The quench-interferometric framework in Fig.\,\ref{fig1} establishes a practical routine to identify QPTs. Based on this framework, we next elaborate that the quantum phase diagram can be obtained by sweeping the quenching parameter $(\kappa, B)$ in the ANNNI Hamiltonian. More specifically, we numerically determines $\kappa_c$ for each fixed $B$ and then varies $B$ from 0 to 1. In Fig.\,\ref{fig3}, we depict the phase diagram obtained from the metrological capability quantified by the QFI, the global quantum fluctuations and the phase estimation uncertainty. 
One can see that the predicted boundary \lxb{all} agrees very well with the theoretical prediction by second-order perturbation theory.

Recently, state-of-the-art quantum simulators have been implemented in a variety of well-controllable many-particle quantum platforms, such as the cold-atom lattice \cite{Bloch2008_RMP, bloch2012quantum, 2017Bloch_ultracoldatoms, schafer2020tools_ultracoldatoms}, the Rydberg-atom array \cite{Adams_2020_Rydbergatom, browaeys2020many, Wu_2021_Rydbergatom}, and the trapped ion crystal \cite{blatt2012quantum, 2021Monroe_trappedions}, which possess long coherence times, tunable interactions and starts to play significant roles in the simulation of a series of interesting spin models. As an example, we point out that the ANNNI chain considered in this work can be effectively synthesized using Rydberg atoms~\cite{2021Das_PRX,zeiher2016many,labuhn2016tunable, Zeiher2017prx,Scholl2022PRXQuantum}. Particularly, the fully polarized initial state can be realized with a high fidelity of $\gtrsim 95\%$~\cite{labuhn2016tunable,Zeiher2017prx, Scholl2022PRXQuantum}. \rev{The variance and the uncertainty of the MLE estimate can be readily obtained by simple global measurement of the collective spin operators \cite{2018RMP_Pezze,Ockeloen2010_pra,muessel2013optimized,Hume2013prl,Dimitrova_2023,colombo2022time,bornet2023scalable,franke2023quantum,Imai2024prl}.} 
Furthermore, the coherent timescale of this system can reach $\tau\geq 10J^{-1}$, where $J$ represents the strength of the nearest-neighbor interaction. Such a long relevant time scale makes the evaluation of the averaged metrological capability over a time window in our probing framework experimentally feasible. 
\lxb{In the Supplementary Materials, we provide a detailed discussion on the operational complexity of the time-averaging requirement \cite{supplement}.}

\vspace*{2ex}
\noindent
\textbf{Conclusion $\&$ outlook}\\
\noindent
To summarize, we present a quench-interferometric framework to probe QPTs by introducing the concept of the metrological capability to characterize the post-quenched state, which exhibit a significant peak around the QCP as quantified by the time average of QFI evolution. This framework can be accomplished through globally measuring collective observables to evaluate the quantum fluctuations or the associated estimation uncertainty of the interferometric phase. By taking the chaotic ANNNI model as a representative example, we demonstrate that the QCPs and immediately the phase diagram can be well identified based on such a framework. The scheme is experimentally feasible even with limited measurement resolution of the observables and under the implementation of non-optimal measurement to extract the phase information encoded in the final state.
\lxb{We also demonstrate that our framework can be extended to other QPT Hamiltonians including long-range interacting models as well \cite{supplement}.}
Our findings take a step of exploiting quantum metrological concepts and strategies towards the investigation of many-body physics, which heuristically unveils a stimulating interplay between quantum metrology and dynamical quantum criticality \cite{2023Chu_prl, 2021Chu_prl, Block2024Scalable}.

\begin{acknowledgments}
\noindent
\textbf{Acknowledgements}\\
\noindent
The computation is completed in the HPC Platform of Huazhong University of Science and Technology.
This work is supported by the National Natural Science Foundation of China (12161141011 and 12174138), the National Key R$\&$D Program of China (2018YFA0306600). Y.-M.C. is also supported by the Young Scientists Fund of the National Natural Science Foundation of China (Grant No. 12304572).
\end{acknowledgments}

\onecolumngrid


\section*{Supplementary material}

# I. The Emergence of the Peak of the QFI at the Phase Transition Point

The Hamiltonian of the ANNNI chain exhibits a $Z_2$ symmetry, i.e. $[U, H(\kappa, B)] = 0$ with $U = \prod_i^N \sigma_i^z$. The initial state that we choose in the main text, i.e. $|\Psi\rangle = |\uparrow\rangle^{\otimes N}$, also satisfies the $Z_2$ symmetry, namely $U|\Psi\rangle = |\Psi\rangle$. Consequently, the time-evolved state $|\Psi_t\rangle = e^{-iH(\kappa,B)t}$ for any time $t$ also satisfies $U|\Psi_t\rangle = |\Psi_t\rangle$. Due to $U^\dagger \hat{S}_x U = -\hat{S}_x$, we have

$$\langle\Psi_t|\hat{S}_x|\Psi_t\rangle = \langle\Psi_t|U^\dagger \hat{S}_x U|\Psi_t\rangle = -\langle\Psi_t|\hat{S}_x|\Psi_t\rangle, \tag{S.1}$$

which directly leads to

$$\langle\Psi_t|\hat{S}_x|\Psi_t\rangle = 0. \tag{S.2}$$

Besides $Z_2$ symmetry, the Hamiltonian and the initial state also satisfy translational symmetry (i.e. for the periodic boundary condition) and reflection symmetry. Therefore, our analysis can be confined to the symmetric eigen subspace corresponding to the eigenvalue $+1$ of these symmetries. Let $\{E_i\}$ and $\{|E_i\rangle\}$ represent the eigenvalues and eigenstates of $H(\kappa, B)$ within such a symmetric subspace. Then, the time-evolved state can be spanned as follows

$$|\Psi_t\rangle = \sum_i e^{-iE_i t} \langle E_i|\Psi\rangle |E_i\rangle. \tag{S.3}$$

The QFI using $|\Psi_t\rangle$ as input of the unitary interferometer characterized by $e^{-i\vartheta \hat{S}_x}$ is given by

$$F_Q(t) = F_Q(|\Psi_t\rangle, \hat{S}_x) \tag{S.4}$$

$$= 4\left[\langle\Psi_t|\hat{S}_x^2|\Psi_t\rangle - \langle\Psi_t|\hat{S}_x|\Psi_t\rangle^2\right] \tag{S.5}$$

$$= 4\langle\Psi_t|\hat{S}_x^2|\Psi_t\rangle \tag{S.6}$$

$$= 4\sum_{i,j} e^{-i(E_i-E_j)t} \langle E_i|\Psi\rangle \langle\Psi|E_j\rangle \langle E_j|\hat{S}_x^2|E_i\rangle. \tag{S.7}$$

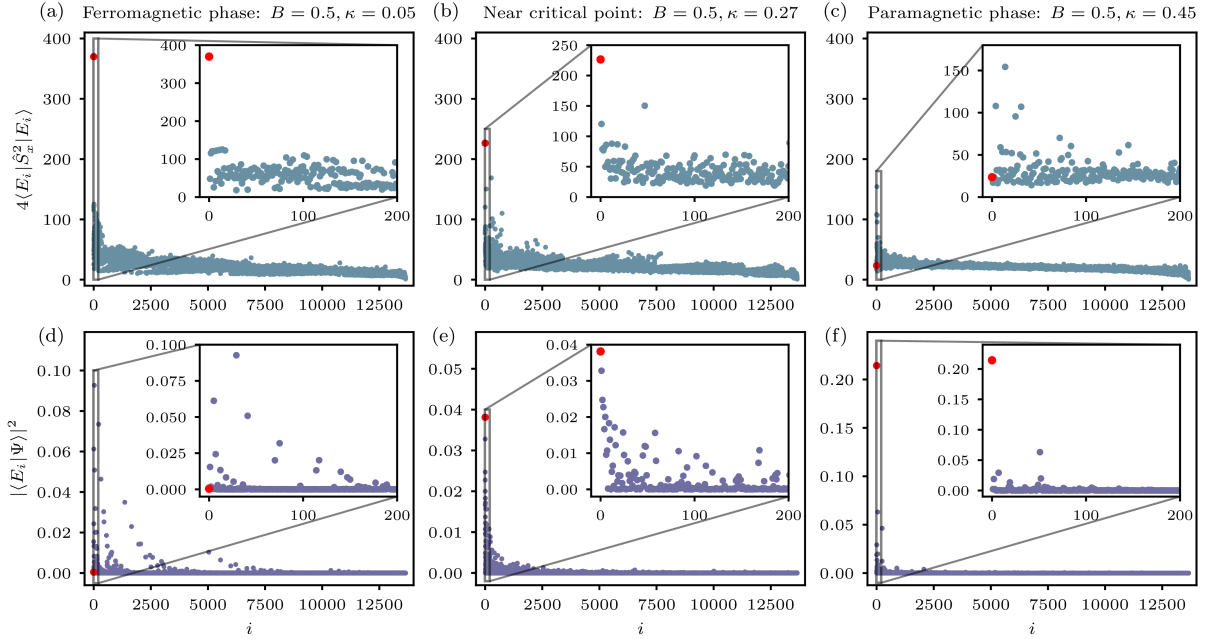

FIG. S1. The time-averaged QFI is determined by the factor $4\langle E_i|\hat{S}_x^2|E_i\rangle$ and the overlap $|\langle E_i|\Psi\rangle|^2$ for all eigenstates within the symmetric subspace. The eigenstates $|E_i\rangle$ are arranged in an ascending order based on the associated eigenenergies $E_i$, with $|E_0\rangle$ denoting the ground state. Results corresponding to the ground state are highlighted with red dots.

Its time-average over a long time window reads,

$$\overline{F_Q} = \lim_{T\to\infty} \frac{1}{T}\int_0^T F_Q(t)dt = 4\sum_i |\langle E_i|\Psi\rangle|^2 \langle E_i|\hat{S}_x^2|E_i\rangle \tag{S.8}$$

We remark that the validity of the above formula requires non-degenerate energy levels. We numerically verify that the ANNNI chain with finite system size and $\kappa \neq 0, B \neq 0$ exhibits no energy degeneracy in the symmetric subspace.

In Fig. S1, we plot $|\langle E_i|\Psi\rangle|^2$ and $\langle E_i|\hat{S}_x^2|E_i\rangle$ of all the eigenstates $|E_i\rangle$ for the ANNNI Hamiltonian in different phases as well as near the QCP, which can be roughly divided into three scenarios: (i) Deep in the ferromagnetic phase, $\langle E_0|\hat{S}_x^2|E_0\rangle$ of the ground state (we denote the ground state as $|E_0\rangle$) is extremely large, as shown in panel (a) of Fig. S1. However, the associated ferromagnetic ground state has very little overlap with the initial state deep in the paramagnetic

3

phase, as shown in Fig. S1 (d). (ii) When the ANNNI Hamiltonian is deep in the paramagnetic phase, there is significant overlap between the initial state and the ground state, as shown in Fig. S1 (f). Nevertheless, $\langle E_0|\hat{S}_x^2|E_0\rangle$ is small in the disordered paramagnetic phase, as shown in S1 (c). (iii) In the vicinity of the critical point, both the factor $\langle E_0|\hat{S}_x^2|E_0\rangle$ and the overlap between the initial state and the ground state are relatively large, resulting in the emergence of the peak of $\overline{F_Q}$, as can be seen in Fig. S1 (b) and (e).

Similar behaviors also occur in other models, including the transverse Ising model, the XY model and the antiferromagnetic Ising chain with both transverse and longitudinal fields. We note that the above analysis only considers the contribution of the ground state. When excited states also has a non-negligible contribution, a more rigorous theoretical analysis deserves intense efforts in future research.

## II. Finite-Size Effect on Locating the QPT with Time-Averaged QFI

To explore the influence of finite-size effect, we calculate the time-averaged QFI for different system sizes. The corresponding results are shown in Fig. S2. It can be seen that our method can effectively locate the QCP even using systems of relatively small sizes. As the system size increase from $N = 8$ to $N = 20$, the location of achieving the maximum $\overline{F_Q}$, denoted as $\kappa = \kappa_{\max}$, move towards the critical point $\kappa_c$ predicted by second-order perturbation theory.

## III. Probing the QPT of the ANNNI Model under Open Boundary Condition

In this section, we consider the quench dynamics of the ANNNI model with open boundary condition. Due to the lack of translational symmetry, we utilize the time-evolving block decimation (TEBD) method [1–3] in this case. We compute the time-averaged QFI for two different system size, $N = 24$ and $N = 30$ spins. We fix the parameter $B$ at $0.25$, $0.5$, $0.75$ and then vary $\kappa$. In order to guarantee the accuracy of the results, we restrict our calculations of the QFI evolution to $T = 30$

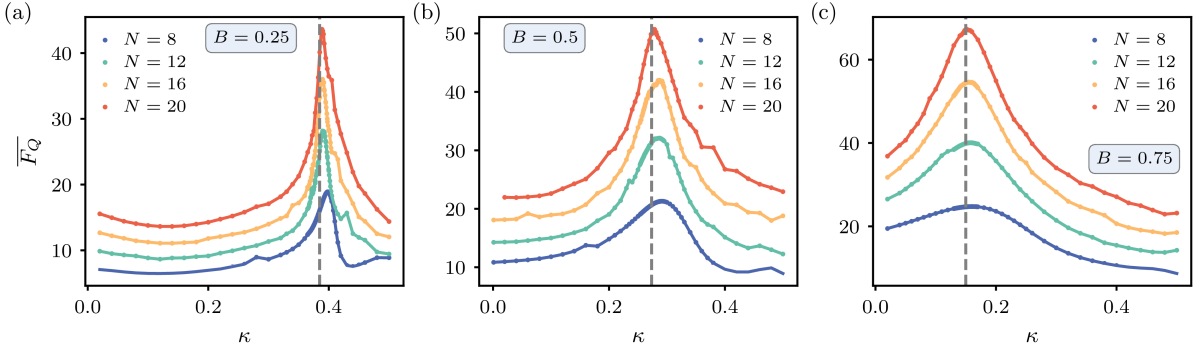

FIG. S2. Time-averaged QFI [see Eq. (S.8)] of ANNNI model plotted as functions of $\kappa$ for different system sizes. We choose the parameter $B$ as 0.25, 0.5, 0.75 in panels (a-c) respectively. The vertical dashed lines indicate the critical point $\kappa_c$ predicted by second-order perturbation theory.

for $N = 24$ and $T = 15$ for $N = 30$. As can be seen in Fig. S3, while the finite size effect is more pronounced under the open boundary condition, the time-averaged QFI still displays prominent peaks near the critical point, implying the robustness of our findings with respect to the open boundary condition.

## IV. Extracting Quantum Fluctuations with Finite Measurement Resolution

In the main text, we propose to repeatedly measure the collective observable $\hat{S}_x = \sum_i \sigma_i^x/2 = \sum_{S_x} S_x \hat{\Pi}_{S_x}$ to extract its quantum fluctuation. Here, $S_x$ represents the eigenvalue of $\hat{S}_x$ and $\hat{\Pi}_{S_x}$ is the corresponding projection operator. To model the imperfect detection with finite resolution, we replace the ideal probabilities $P(S_x) = \langle \Psi_t | \hat{\Pi}_{S_x} | \Psi_t \rangle$ with a smeared probability density [5]

$$P_{\mathrm{dn}}(S_x') = \sum_{S_x} P(S_x'|S_x) P(S_x), \tag{S.9}$$

where $P(S_x'|S_x)$ is a convolution function representing the probability to obtain the result $S_x'$ when the actual value is $S_x$. Typically, this function is modeled as a normalized Gaussian $P(S_x'|S_x) = e^{-(S_x'-S_x)^2/(2\Delta^2)}/(\sqrt{2\pi}\Delta)$, where the parameter $\Delta$

5

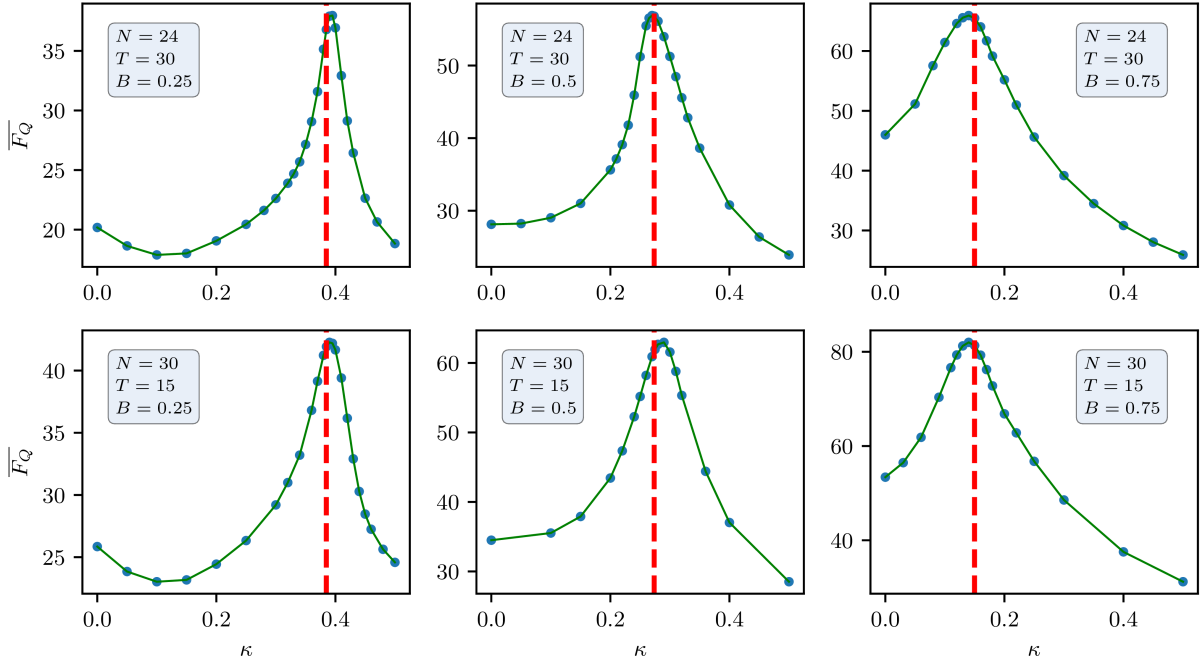

FIG. S3. TEBD results for ANNNI chain under open boundary conditions with $N = 24$ and $N = 30$ spins. The time-averaged QFI is plotted as functions of $\kappa$ with $B$ fixed at $0.25$, $0.5$, and $0.75$. We note the time-averaging window is $T = 30$ for $N = 24$ and $T = 15$ for $N = 30$. The vertical dashed lines mark the critical point $\kappa_c$ of the corresponding $B$ predicted by second-order perturbation theory [4].

quantifies the measurement resolution. This approach aligns with the model for coarse-grained detectors described in Ref. [6], which assumes a continuous measurement

$$\hat{\Pi}(S'_x) = \frac{1}{\sqrt{2\pi}\Delta} e^{-(S'_x - \hat{S}_x)^2/(2\Delta^2)} = \sum_{S_x} \frac{1}{\sqrt{2\pi}\Delta} e^{-(S'_x - S_x)^2/(2\Delta^2)} \hat{\Pi}_{S_x}. \quad \text{(S.10)}$$

This measurement return to the perfect measurement $\hat{\Pi}(S'_x) = \delta(S'_x - \hat{S}_x)$ for $\Delta \to 0$.

**V. Probing the QPT with Parameter Estimation Uncertainty**

In the main text, we consider the collective measurement, $\hat{S}_z = \sum_i \sigma^z_i / 2 = \sum_{S_z} S_z \hat{\Pi}_{S_z}$, to estimate the unknown phase shift $\vartheta$ encoded in the multi-spin state

6

$|\Psi_t(\vartheta)\rangle$, where $S_z$ denotes the eigenvalue of $\hat{S}_z$ and $\hat{\Pi}_{S_z}$ is the corresponding projection operator. Firstly, we demonstrate that the Fisher information extracted by this simple global measurement can reach the QFI as $\vartheta \to 0$ and exhibits nearly synchronous dynamics as the QFI when $\vartheta$ is small. Specifically, the Fisher information associated with the measurement of $\hat{S}_z$ is given by

$$F_C = \sum_{S_z} P(S_z|\vartheta) \left(\frac{\partial \ln P(S_z|\vartheta)}{\partial \vartheta}\right)^2, \qquad (S.11)$$

where $P(S_z|\vartheta) = \langle \Psi_t(\vartheta)|\hat{\Pi}_{S_z}|\Psi_t(\vartheta)\rangle$ is the probability of obtaining the outcome of $S_z$. Generally, $F_C$ depends on the value of the parameter $\vartheta$. In Fig. S4 (a), we plot how $F_C$ varies with $\vartheta$ for $B = 0.75$, $\kappa = 0.15$ and $t = 7.5$. We observe that $F_C$ is significantly enhanced and approaches the QFI ($F_Q$) when $\vartheta = n\pi$ with $n$ taking the value of integers. In Fig. S4 (b), we present the time evolution of $F_C$ for different values of $\vartheta$ and compare these results to $F_Q$. The results show that when $\vartheta$ is small, $F_C$ is very close to $F_Q$ and show strongly correlated behaviors. Moreover,

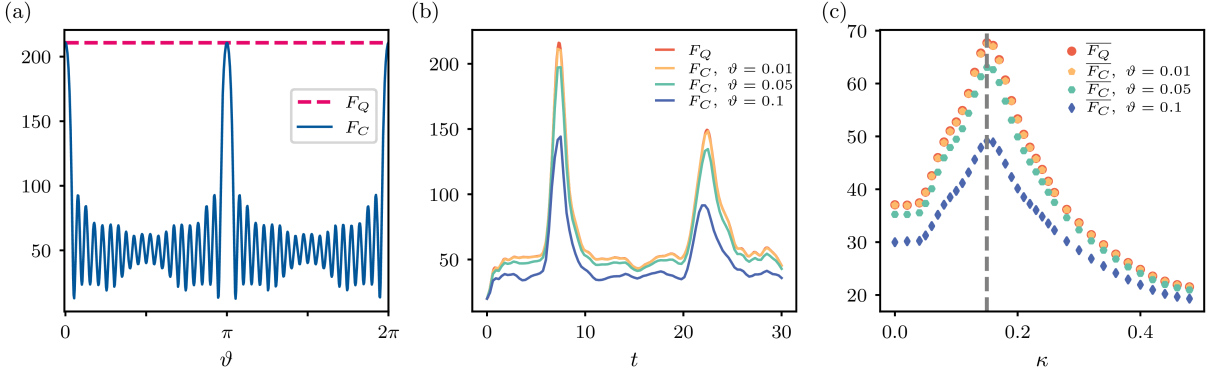

FIG. S4. (a) The Fisher information associated with the measurement of $\hat{S}_z$ as a function of the phase $\vartheta$ at the evolution time $t = 7.5$. (b) Time evolution of $F_C$ for different values of $\vartheta$. (c) Time-averaged QFI ($\overline{F_Q}$) and FI ($\overline{F_C}$) plotted as functions of $\kappa$ over a time window of $T = 15$. The vertical dashed line indicates the critical point of $\kappa_c \approx 0.15$. Here, the system parameters of the ANNNI model are set as $\kappa = \kappa_c = 0.15$ in panels (a-b) and $B = 0.75$ in all panels. All the calculations are performed for a system size of $N = 20$.

7

we calculate the time-averaged QFI and FI over a time window of $T = 15$, denoted by $\overline{F_Q}$ and $\overline{F_C}$ respectively. The results are shown in Fig. S4 (c). It is evident that $\overline{F_C}$ also exhibits distinct peaks near the critical point, indicating that the QPT can be successfully detected by non-optimized metrological measurement to extract the phase $\vartheta$.

In order to unbiasedly estimate the phase $\vartheta$ from the measurement outcomes, we consider the approach of maximum likelihood estimation (MLE). Assuming that the outcomes on $n$ copies of the parameterized quantum state $|\Psi_t(\vartheta)\rangle$ as $\{S_z^{(i)}\}_{i=1}^n = \{S_z^{(1)}, S_z^{(2)}, \cdots, S_z^{(n)}\}$, the log-likelihood function is then defined as

$$l\left(\vartheta; \{S_z^{(i)}\}_{i=1}^n\right) = \sum_{i=1}^n \ln P(S_z^{(i)}|\vartheta), \tag{S.12}$$

where $P(S_z^{(i)}|\vartheta)$ denotes the probability of obtaining the outcome $S_z^{(i)}$. We numerically calculate the log-likelihood function $l\left(\vartheta; \{S_{zi}\}_{i=1}^n\right)$ over a range that covers the possible values of the unknown phase $\vartheta$ and find its maximum point within this range. MLE estimate corresponds to the value of $\vartheta$ that achieves the maximum of the log-likelihood function [7].

**VI. Dynamical quench-interferometric framework for other QPT models**

*(A) Probing the QPT of the transverse field Ising chain.—* The transverse field Ising chain is one of the most studied models in quantum phase transitions [8, 9]. Its Hamiltonian is given by

$$H(B) = -\sum_i \sigma_i^x \sigma_{i+1}^x - B \sum_i \sigma_i^z. \tag{S.13}$$

This model represents the noninteracting limit ($\kappa = 0$) of the ANNNI Hamiltonian [see Eq. (3) of the main text)] and it exhibits a quantum phase transition at $B_c = 1$ from a ferromagnetic phase ($B < 1$) to a paramagnetic phase ($B > 1$). We propose to prepare the spin chain in a fully polarized initial state $|\Psi\rangle = |\uparrow\rangle^{\otimes N}$,

and let it evolve under the quenched Hamiltonian $H(B)$. The time-averaged QFI $\overline{F_Q} = \int_0^T F_Q(|\Psi_t\rangle, \hat{S}_x)\,\mathrm{d}t$, calculated over a time window of $T = 30$, is plotted as a function of the quench parameter $B$ in Fig. S5. It can be seen that the $\overline{F_Q}$ is considerably enhanced near the critical point.

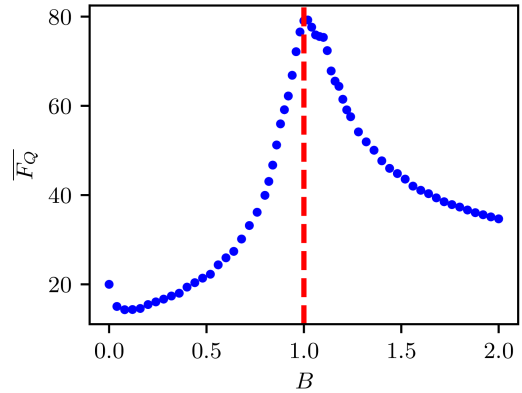

FIG. S5. Time-averaged QFI for the transverse Ising chain. The vertical dashed lines mark the critical point $B_c = 1$. Here the system size is set to $N = 20$, and the time-averaging window is $T = 30$.

*(B) Probing the QPT of the XY chain.—* The Hamiltonian of the XY chain [10] is

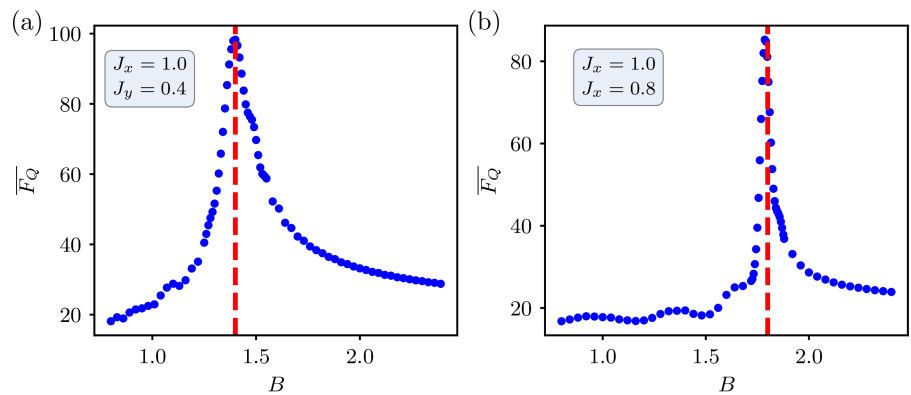

FIG. S6. Time-averaged QFI for the XY chain. Here the system size is set to $N = 20$, and the averaging time window is $T = 30$. The vertical dashed lines mark the critical points at $B_c = 1.4$ for (a) and $B_c = 1.8$ for (b) respectively.

9

given by

$$H(B) = -J_x \sum_i \sigma_i^x \sigma_{i+1}^x - J_y \sum_i \sigma_i^y \sigma_{i+1}^y - B \sum_i \sigma_i^z. \quad \text{(S.14)}$$

Without loss of generality, we assume that $J_x \geq |J_y| \geq 0$ (thus $J_x + J_y > 0$). The system undergoes transitions from paramagnetic phase to ferromagnetic phase at $B = \pm(J_x + J_y)$. We assume to initialize the spin chain in a fully polarized state $|\Psi\rangle = |\uparrow\rangle^{\otimes N}$ and let it evolve under the quenched Hamiltonian $H(B)$. The time-averaged QFI $\overline{F_Q} = \int_0^T F_Q(|\Psi_t\rangle, \hat{S}_x) \, dt$, calculated over a time window of $T = 30$, is plotted as a function of the quench parameter $B$ in Fig. S5. We choose $J_x = 1, J_y = 0.4$ for Fig. S6 (a) and $J_x = 1, J_y = 0.8$ for Fig. S6 (b). It is evident that $\overline{F_Q}$ is significantly enhanced near the critical point.

*(C) Probing QPT of the Ising chain with transverse and longitudinal fields.—* The Hamiltonian of the antiferromagnetic Ising chain with both transverse and longitudinal fields [11–13] is written as

$$H(B_x, B_z) = \sum_i \sigma_i^z \sigma_{i+1}^z - B_x \sum_i \sigma_i^x - B_z \sum_i \sigma_i^z. \quad \text{(S.15)}$$

This is a nonintegrable model and exhibits two phases at zero temperature [11, 12], namely an antiferromagnetic phase for weak fields, and a paramagnetic phase for strong fields. These phases are separated by a second-order phase transition [see the dashed line in Fig. S7 (b)].

Similarly, we assume that the spin chain is initialized in a polarized state $|\Psi\rangle = |\rightarrow\rangle^{\otimes N}$ and consider the evolution governed by $H(B_x, B_z)$. The time-averaged QFI, $\overline{F_Q} = \int_0^T F_Q(|\Psi_t\rangle, \hat{\mathcal{O}}) \, dt$, is computed by choosing the generator of the unitary interferometer as $\hat{\mathcal{O}} = \sum_i (-1)^i \sigma_i^z / 2$. To probe the antiferromagnetic to paramagnetic quantum phase transition at a fixed value of $B_x$, we fix the quench parameter $B_x$ and vary $B_z$. For example, we show the results for $B_x$ fixed at $0.437$ in Fig. S7, where the vertical dashed line marks the critical point, $(B_z)_c$, from Ref. [11] obtained by

10

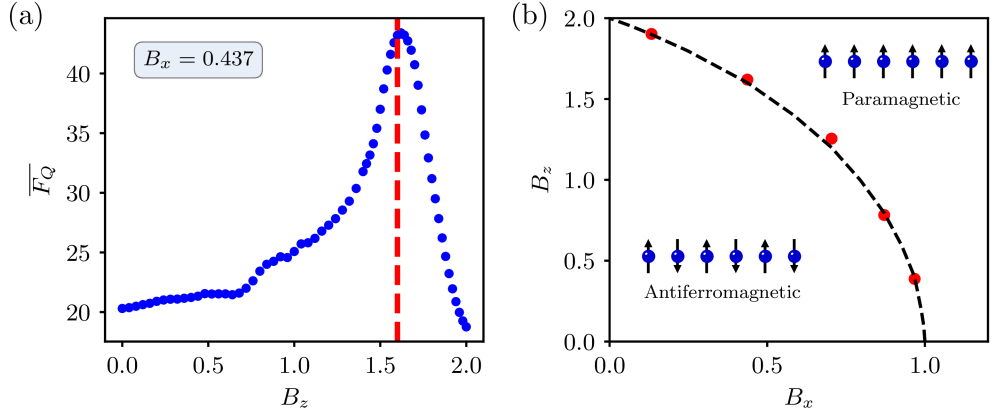

FIG. S7. (a) Time-averaged QFI plotted as a function of $B_z$ when fixing $B_x = 0.437$. The vertical dashed line marks the critical point, $(B_z)_c$, predicted from the DMRG calculation. (b) Phase diagram in the parameter plane of $(B_x, B_z)$. The phase boundary is well described by locating the peak of the time-averaged QFI over a window of $T = 30$ (red dots). The dashed line is the critical line from Ref. [11] calculated using DMRG. Here the system size is set as $N = 20$.

the DMRG calculation. It can be seen that the position where $\overline{F_Q}$ reaches the maximum coincides with the critical $(B_z)_c$. In Fig. S7 (b), we further present the critical points estimated from the maximum of the time-averaged $\overline{F_Q}$ for various $B_x$. These results are in good agreement with the phase boundaries determined by the DMRG calculation.

---

* yaomingchu@hust.edu.cn

† jianmingcai@hust.edu.cn

[1] G. Vidal, Efficient Classical Simulation of Slightly Entangled Quantum Computations, Phys. Rev. Lett. **91**, 147902 (2003).

[2] G. Vidal, Efficient Simulation of One-Dimensional Quantum Many-Body Systems, Phys. Rev. Lett. **93**, 040502 (2004).

[3] J. Hauschild and F. Pollmann, Efficient numerical simulations with Tensor Networks:

11




\begin{thebibliography}{98}%
\makeatletter
\providecommand \@ifxundefined [1]{%
 \@ifx{#1\undefined}
}%
\providecommand \@ifnum [1]{%
 \ifnum #1\expandafter \@firstoftwo
 \else \expandafter \@secondoftwo
 \fi
}%
\providecommand \@ifx [1]{%
 \ifx #1\expandafter \@firstoftwo
 \else \expandafter \@secondoftwo
 \fi
}%
\providecommand \natexlab [1]{#1}%
\providecommand \enquote  [1]{``#1''}%
\providecommand \bibnamefont  [1]{#1}%
\providecommand \bibfnamefont [1]{#1}%
\providecommand \citenamefont [1]{#1}%
\providecommand \href@noop [0]{\@secondoftwo}%
\providecommand \href [0]{\begingroup \@sanitize@url \@href}%
\providecommand \@href[1]{\@@startlink{#1}\@@href}%
\providecommand \@@href[1]{\endgroup#1\@@endlink}%
\providecommand \@sanitize@url [0]{\catcode `\\12\catcode `\$12\catcode
  `\&12\catcode `\#12\catcode `\^12\catcode `\_12\catcode `\%12\relax}%
\providecommand \@@startlink[1]{}%
\providecommand \@@endlink[0]{}%
\providecommand \url  [0]{\begingroup\@sanitize@url \@url }%
\providecommand \@url [1]{\endgroup\@href {#1}{\urlprefix }}%
\providecommand \urlprefix  [0]{URL }%
\providecommand \Eprint [0]{\href }%
\providecommand \doibase [0]{https://doi.org/}%
\providecommand \selectlanguage [0]{\@gobble}%
\providecommand \bibinfo  [0]{\@secondoftwo}%
\providecommand \bibfield  [0]{\@secondoftwo}%
\providecommand \translation [1]{[#1]}%
\providecommand \BibitemOpen [0]{}%
\providecommand \bibitemStop [0]{}%
\providecommand \bibitemNoStop [0]{.\EOS\space}%
\providecommand \EOS [0]{\spacefactor3000\relax}%
\providecommand \BibitemShut  [1]{\csname bibitem#1\endcsname}%
\let\auto@bib@innerbib\@empty
\bibitem [{\citenamefont {Sachdev}(2011)}]{sachdev2011quantum}%
  \BibitemOpen
  \bibfield  {author} {\bibinfo {author} {\bibfnamefont {S.}~\bibnamefont
  {Sachdev}},\ }\href@noop {} {\emph {\bibinfo {title} {{Quantum Phase
  Transitions}}}}\ (\bibinfo  {publisher} {Cambridge University Press},\
  \bibinfo {year} {2011})\BibitemShut {NoStop}%
\bibitem [{\citenamefont {Vojta}(2003)}]{MatthiasVojta_2003}%
  \BibitemOpen
  \bibfield  {author} {\bibinfo {author} {\bibfnamefont {M.}~\bibnamefont
  {Vojta}},\ }\bibfield  {title} {\bibinfo {title} {{Quantum phase
  transitions}},\ }\href {https://doi.org/10.1088/0034-4885/66/12/R01}
  {\bibfield  {journal} {\bibinfo  {journal} {Reports on Progress in Physics}\
  }\textbf {\bibinfo {volume} {66}},\ \bibinfo {pages} {2069} (\bibinfo {year}
  {2003})}\BibitemShut {NoStop}%
\bibitem [{\citenamefont {Sachdev}(2008)}]{sachdev2008quantum}%
  \BibitemOpen
  \bibfield  {author} {\bibinfo {author} {\bibfnamefont {S.}~\bibnamefont
  {Sachdev}},\ }\bibfield  {title} {\bibinfo {title} {{Quantum magnetism and
  criticality}},\ }\href {https://doi.org/10.1038/nphys894} {\bibfield
  {journal} {\bibinfo  {journal} {Nature Physics}\ }\textbf {\bibinfo {volume}
  {4}},\ \bibinfo {pages} {173} (\bibinfo {year} {2008})}\BibitemShut {NoStop}%
\bibitem [{\citenamefont {Savary}\ and\ \citenamefont
  {Balents}(2016)}]{Savary_2017}%
  \BibitemOpen
  \bibfield  {author} {\bibinfo {author} {\bibfnamefont {L.}~\bibnamefont
  {Savary}}\ and\ \bibinfo {author} {\bibfnamefont {L.}~\bibnamefont
  {Balents}},\ }\bibfield  {title} {\bibinfo {title} {{Quantum spin liquids: a
  review}},\ }\href {https://doi.org/10.1088/0034-4885/80/1/016502} {\bibfield
  {journal} {\bibinfo  {journal} {Reports on Progress in Physics}\ }\textbf
  {\bibinfo {volume} {80}},\ \bibinfo {pages} {016502} (\bibinfo {year}
  {2016})}\BibitemShut {NoStop}%
\bibitem [{\citenamefont {Wen}(2013)}]{Wen2013_topological_order}%
  \BibitemOpen
  \bibfield  {author} {\bibinfo {author} {\bibfnamefont {X.-G.}\ \bibnamefont
  {Wen}},\ }\bibfield  {title} {\bibinfo {title} {{Topological Order: From
  Long-Range Entangled Quantum Matter to a Unified Origin of Light and
  Electrons}},\ }\href {https://doi.org/https://doi.org/10.1155/2013/198710}
  {\bibfield  {journal} {\bibinfo  {journal} {International Scholarly Research
  Notices}\ }\textbf {\bibinfo {volume} {2013}},\ \bibinfo {pages} {198710}
  (\bibinfo {year} {2013})}\BibitemShut {NoStop}%
\bibitem [{\citenamefont {Gantmakher}\ and\ \citenamefont
  {Dolgopolov}(2010)}]{Gantmakher_2010}%
  \BibitemOpen
  \bibfield  {author} {\bibinfo {author} {\bibfnamefont {V.~F.}\ \bibnamefont
  {Gantmakher}}\ and\ \bibinfo {author} {\bibfnamefont {V.~T.}\ \bibnamefont
  {Dolgopolov}},\ }\bibfield  {title} {\bibinfo {title}
  {{Superconductor–insulator quantum phase transition}},\ }\href
  {https://doi.org/10.3367/UFNe.0180.201001a.0003} {\bibfield  {journal}
  {\bibinfo  {journal} {Physics-Uspekhi}\ }\textbf {\bibinfo {volume} {53}},\
  \bibinfo {pages} {1} (\bibinfo {year} {2010})}\BibitemShut {NoStop}%
\bibitem [{\citenamefont {Zhang}\ \emph {et~al.}(2017)\citenamefont {Zhang},
  \citenamefont {Pagano}, \citenamefont {Hess}, \citenamefont {Kyprianidis},
  \citenamefont {Becker}, \citenamefont {Kaplan}, \citenamefont {Gorshkov},
  \citenamefont {Gong},\ and\ \citenamefont {Monroe}}]{zhang2017observation}%
  \BibitemOpen
  \bibfield  {author} {\bibinfo {author} {\bibfnamefont {J.}~\bibnamefont
  {Zhang}}, \bibinfo {author} {\bibfnamefont {G.}~\bibnamefont {Pagano}},
  \bibinfo {author} {\bibfnamefont {P.~W.}\ \bibnamefont {Hess}}, \bibinfo
  {author} {\bibfnamefont {A.}~\bibnamefont {Kyprianidis}}, \bibinfo {author}
  {\bibfnamefont {P.}~\bibnamefont {Becker}}, \bibinfo {author} {\bibfnamefont
  {H.}~\bibnamefont {Kaplan}}, \bibinfo {author} {\bibfnamefont {A.~V.}\
  \bibnamefont {Gorshkov}}, \bibinfo {author} {\bibfnamefont {Z.-X.}\
  \bibnamefont {Gong}},\ and\ \bibinfo {author} {\bibfnamefont
  {C.}~\bibnamefont {Monroe}},\ }\bibfield  {title} {\bibinfo {title}
  {{Observation of a many-body dynamical phase transition with a 53-qubit
  quantum simulator}},\ }\href {https://doi.org/10.1038/nature24654} {\bibfield
   {journal} {\bibinfo  {journal} {Nature}\ }\textbf {\bibinfo {volume}
  {551}},\ \bibinfo {pages} {601} (\bibinfo {year} {2017})}\BibitemShut
  {NoStop}%
\bibitem [{\citenamefont {Georgescu}\ \emph {et~al.}(2014)\citenamefont
  {Georgescu}, \citenamefont {Ashhab},\ and\ \citenamefont
  {Nori}}]{Georgescu2014_RMP}%
  \BibitemOpen
  \bibfield  {author} {\bibinfo {author} {\bibfnamefont {I.~M.}\ \bibnamefont
  {Georgescu}}, \bibinfo {author} {\bibfnamefont {S.}~\bibnamefont {Ashhab}},\
  and\ \bibinfo {author} {\bibfnamefont {F.}~\bibnamefont {Nori}},\ }\bibfield
  {title} {\bibinfo {title} {{Quantum simulation}},\ }\href
  {https://doi.org/10.1103/RevModPhys.86.153} {\bibfield  {journal} {\bibinfo
  {journal} {Rev. Mod. Phys.}\ }\textbf {\bibinfo {volume} {86}},\ \bibinfo
  {pages} {153} (\bibinfo {year} {2014})}\BibitemShut {NoStop}%
\bibitem [{\citenamefont {Lv}\ \emph {et~al.}(2018)\citenamefont {Lv},
  \citenamefont {An}, \citenamefont {Liu}, \citenamefont {Zhang}, \citenamefont
  {Pedernales}, \citenamefont {Lamata}, \citenamefont {Solano},\ and\
  \citenamefont {Kim}}]{Dingshun2018prx}%
  \BibitemOpen
  \bibfield  {author} {\bibinfo {author} {\bibfnamefont {D.}~\bibnamefont
  {Lv}}, \bibinfo {author} {\bibfnamefont {S.}~\bibnamefont {An}}, \bibinfo
  {author} {\bibfnamefont {Z.}~\bibnamefont {Liu}}, \bibinfo {author}
  {\bibfnamefont {J.-N.}\ \bibnamefont {Zhang}}, \bibinfo {author}
  {\bibfnamefont {J.~S.}\ \bibnamefont {Pedernales}}, \bibinfo {author}
  {\bibfnamefont {L.}~\bibnamefont {Lamata}}, \bibinfo {author} {\bibfnamefont
  {E.}~\bibnamefont {Solano}},\ and\ \bibinfo {author} {\bibfnamefont
  {K.}~\bibnamefont {Kim}},\ }\bibfield  {title} {\bibinfo {title} {{Quantum
  Simulation of the Quantum Rabi Model in a Trapped Ion}},\ }\href
  {https://doi.org/10.1103/PhysRevX.8.021027} {\bibfield  {journal} {\bibinfo
  {journal} {Phys. Rev. X}\ }\textbf {\bibinfo {volume} {8}},\ \bibinfo {pages}
  {021027} (\bibinfo {year} {2018})}\BibitemShut {NoStop}%
\bibitem [{\citenamefont {Harris}\ \emph {et~al.}(2018)\citenamefont {Harris},
  \citenamefont {Sato}, \citenamefont {Berkley}, \citenamefont {Reis},
  \citenamefont {Altomare}, \citenamefont {Amin}, \citenamefont {Boothby},
  \citenamefont {Bunyk}, \citenamefont {Deng}, \citenamefont {Enderud},
  \citenamefont {Huang}, \citenamefont {Hoskinson}, \citenamefont {Johnson},
  \citenamefont {Ladizinsky}, \citenamefont {Ladizinsky}, \citenamefont
  {Lanting}, \citenamefont {Li}, \citenamefont {Medina}, \citenamefont
  {Molavi}, \citenamefont {Neufeld}, \citenamefont {Oh}, \citenamefont
  {Pavlov}, \citenamefont {Perminov}, \citenamefont {Poulin-Lamarre},
  \citenamefont {Rich}, \citenamefont {Smirnov}, \citenamefont {Swenson},
  \citenamefont {Tsai}, \citenamefont {Volkmann}, \citenamefont {Whittaker},\
  and\ \citenamefont {Yao}}]{Harris2018science}%
  \BibitemOpen
  \bibfield  {author} {\bibinfo {author} {\bibfnamefont {R.}~\bibnamefont
  {Harris}}, \bibinfo {author} {\bibfnamefont {Y.}~\bibnamefont {Sato}},
  \bibinfo {author} {\bibfnamefont {A.~J.}\ \bibnamefont {Berkley}}, \bibinfo
  {author} {\bibfnamefont {M.}~\bibnamefont {Reis}}, \bibinfo {author}
  {\bibfnamefont {F.}~\bibnamefont {Altomare}}, \bibinfo {author}
  {\bibfnamefont {M.~H.}\ \bibnamefont {Amin}}, \bibinfo {author}
  {\bibfnamefont {K.}~\bibnamefont {Boothby}}, \bibinfo {author} {\bibfnamefont
  {P.}~\bibnamefont {Bunyk}}, \bibinfo {author} {\bibfnamefont
  {C.}~\bibnamefont {Deng}}, \bibinfo {author} {\bibfnamefont {C.}~\bibnamefont
  {Enderud}}, \bibinfo {author} {\bibfnamefont {S.}~\bibnamefont {Huang}},
  \bibinfo {author} {\bibfnamefont {E.}~\bibnamefont {Hoskinson}}, \bibinfo
  {author} {\bibfnamefont {M.~W.}\ \bibnamefont {Johnson}}, \bibinfo {author}
  {\bibfnamefont {E.}~\bibnamefont {Ladizinsky}}, \bibinfo {author}
  {\bibfnamefont {N.}~\bibnamefont {Ladizinsky}}, \bibinfo {author}
  {\bibfnamefont {T.}~\bibnamefont {Lanting}}, \bibinfo {author} {\bibfnamefont
  {R.}~\bibnamefont {Li}}, \bibinfo {author} {\bibfnamefont {T.}~\bibnamefont
  {Medina}}, \bibinfo {author} {\bibfnamefont {R.}~\bibnamefont {Molavi}},
  \bibinfo {author} {\bibfnamefont {R.}~\bibnamefont {Neufeld}}, \bibinfo
  {author} {\bibfnamefont {T.}~\bibnamefont {Oh}}, \bibinfo {author}
  {\bibfnamefont {I.}~\bibnamefont {Pavlov}}, \bibinfo {author} {\bibfnamefont
  {I.}~\bibnamefont {Perminov}}, \bibinfo {author} {\bibfnamefont
  {G.}~\bibnamefont {Poulin-Lamarre}}, \bibinfo {author} {\bibfnamefont
  {C.}~\bibnamefont {Rich}}, \bibinfo {author} {\bibfnamefont {A.}~\bibnamefont
  {Smirnov}}, \bibinfo {author} {\bibfnamefont {L.}~\bibnamefont {Swenson}},
  \bibinfo {author} {\bibfnamefont {N.}~\bibnamefont {Tsai}}, \bibinfo {author}
  {\bibfnamefont {M.}~\bibnamefont {Volkmann}}, \bibinfo {author}
  {\bibfnamefont {J.}~\bibnamefont {Whittaker}},\ and\ \bibinfo {author}
  {\bibfnamefont {J.}~\bibnamefont {Yao}},\ }\bibfield  {title} {\bibinfo
  {title} {Phase transitions in a programmable quantum spin glass simulator},\
  }\href {https://doi.org/10.1126/science.aat2025} {\bibfield  {journal}
  {\bibinfo  {journal} {Science}\ }\textbf {\bibinfo {volume} {361}},\ \bibinfo
  {pages} {162} (\bibinfo {year} {2018})}\BibitemShut {NoStop}%
\bibitem [{\citenamefont {Daley}\ \emph {et~al.}(2022)\citenamefont {Daley},
  \citenamefont {Bloch}, \citenamefont {Kokail}, \citenamefont {Flannigan},
  \citenamefont {Pearson}, \citenamefont {Troyer},\ and\ \citenamefont
  {Zoller}}]{Daley2022quantumsimulation}%
  \BibitemOpen
  \bibfield  {author} {\bibinfo {author} {\bibfnamefont {A.~J.}\ \bibnamefont
  {Daley}}, \bibinfo {author} {\bibfnamefont {I.}~\bibnamefont {Bloch}},
  \bibinfo {author} {\bibfnamefont {C.}~\bibnamefont {Kokail}}, \bibinfo
  {author} {\bibfnamefont {S.}~\bibnamefont {Flannigan}}, \bibinfo {author}
  {\bibfnamefont {N.}~\bibnamefont {Pearson}}, \bibinfo {author} {\bibfnamefont
  {M.}~\bibnamefont {Troyer}},\ and\ \bibinfo {author} {\bibfnamefont
  {P.}~\bibnamefont {Zoller}},\ }\bibfield  {title} {\bibinfo {title}
  {Practical quantum advantage in quantum simulation},\ }\href
  {https://doi.org/10.1038/s41586-022-04940-6} {\bibfield  {journal} {\bibinfo
  {journal} {Nature}\ }\textbf {\bibinfo {volume} {607}},\ \bibinfo {pages}
  {667} (\bibinfo {year} {2022})}\BibitemShut {NoStop}%
\bibitem [{\citenamefont {Guo}\ \emph {et~al.}(2024)\citenamefont {Guo},
  \citenamefont {Wu}, \citenamefont {Ye}, \citenamefont {Zhang}, \citenamefont
  {Lian}, \citenamefont {Yao}, \citenamefont {Wang}, \citenamefont {Yan},
  \citenamefont {Yi}, \citenamefont {Xu} \emph {et~al.}}]{Duan2024trappedions}%
  \BibitemOpen
  \bibfield  {author} {\bibinfo {author} {\bibfnamefont {S.-A.}\ \bibnamefont
  {Guo}}, \bibinfo {author} {\bibfnamefont {Y.-K.}\ \bibnamefont {Wu}},
  \bibinfo {author} {\bibfnamefont {J.}~\bibnamefont {Ye}}, \bibinfo {author}
  {\bibfnamefont {L.}~\bibnamefont {Zhang}}, \bibinfo {author} {\bibfnamefont
  {W.-Q.}\ \bibnamefont {Lian}}, \bibinfo {author} {\bibfnamefont
  {R.}~\bibnamefont {Yao}}, \bibinfo {author} {\bibfnamefont {Y.}~\bibnamefont
  {Wang}}, \bibinfo {author} {\bibfnamefont {R.-Y.}\ \bibnamefont {Yan}},
  \bibinfo {author} {\bibfnamefont {Y.-J.}\ \bibnamefont {Yi}}, \bibinfo
  {author} {\bibfnamefont {Y.-L.}\ \bibnamefont {Xu}}, \emph {et~al.},\
  }\bibfield  {title} {\bibinfo {title} {A site-resolved two-dimensional
  quantum simulator with hundreds of trapped ions},\ }\href
  {https://doi.org/10.1038/s41586-024-07459-0} {\bibfield  {journal} {\bibinfo
  {journal} {Nature}\ }\textbf {\bibinfo {volume} {630}},\ \bibinfo {pages}
  {613} (\bibinfo {year} {2024})}\BibitemShut {NoStop}%
\bibitem [{\citenamefont {Shao}\ \emph {et~al.}(2024)\citenamefont {Shao},
  \citenamefont {Wang}, \citenamefont {Zhu}, \citenamefont {Zhu}, \citenamefont
  {Sun}, \citenamefont {Chen}, \citenamefont {Zhang}, \citenamefont {Fan},
  \citenamefont {Deng}, \citenamefont {Yao} \emph
  {et~al.}}]{Pan2024_antiferromagnetic}%
  \BibitemOpen
  \bibfield  {author} {\bibinfo {author} {\bibfnamefont {H.-J.}\ \bibnamefont
  {Shao}}, \bibinfo {author} {\bibfnamefont {Y.-X.}\ \bibnamefont {Wang}},
  \bibinfo {author} {\bibfnamefont {D.-Z.}\ \bibnamefont {Zhu}}, \bibinfo
  {author} {\bibfnamefont {Y.-S.}\ \bibnamefont {Zhu}}, \bibinfo {author}
  {\bibfnamefont {H.-N.}\ \bibnamefont {Sun}}, \bibinfo {author} {\bibfnamefont
  {S.-Y.}\ \bibnamefont {Chen}}, \bibinfo {author} {\bibfnamefont
  {C.}~\bibnamefont {Zhang}}, \bibinfo {author} {\bibfnamefont {Z.-J.}\
  \bibnamefont {Fan}}, \bibinfo {author} {\bibfnamefont {Y.}~\bibnamefont
  {Deng}}, \bibinfo {author} {\bibfnamefont {X.-C.}\ \bibnamefont {Yao}}, \emph
  {et~al.},\ }\bibfield  {title} {\bibinfo {title} {{Antiferromagnetic phase
  transition in a 3D fermionic Hubbard model}},\ }\href
  {https://doi.org/10.1038/s41586-024-07689-2} {\bibfield  {journal} {\bibinfo
  {journal} {Nature}\ ,\ \bibinfo {pages} {1}} (\bibinfo {year}
  {2024})}\BibitemShut {NoStop}%
\bibitem [{\citenamefont {Osterloh}\ \emph {et~al.}(2002)\citenamefont
  {Osterloh}, \citenamefont {Amico}, \citenamefont {Falci},\ and\ \citenamefont
  {Fazio}}]{osterloh2002scaling}%
  \BibitemOpen
  \bibfield  {author} {\bibinfo {author} {\bibfnamefont {A.}~\bibnamefont
  {Osterloh}}, \bibinfo {author} {\bibfnamefont {L.}~\bibnamefont {Amico}},
  \bibinfo {author} {\bibfnamefont {G.}~\bibnamefont {Falci}},\ and\ \bibinfo
  {author} {\bibfnamefont {R.}~\bibnamefont {Fazio}},\ }\bibfield  {title}
  {\bibinfo {title} {{Scaling of entanglement close to a quantum phase
  transition}},\ }\href {https://doi.org/10.1038/416608a} {\bibfield  {journal}
  {\bibinfo  {journal} {Nature}\ }\textbf {\bibinfo {volume} {416}},\ \bibinfo
  {pages} {608} (\bibinfo {year} {2002})}\BibitemShut {NoStop}%
\bibitem [{\citenamefont {Gu}(2010)}]{Gu2010}%
  \BibitemOpen
  \bibfield  {author} {\bibinfo {author} {\bibfnamefont {S.-J.}\ \bibnamefont
  {Gu}},\ }\bibfield  {title} {\bibinfo {title} {Fidelity approach to quantum
  phase transitions},\ }\href {https://doi.org/10.1142/S0217979210056335}
  {\bibfield  {journal} {\bibinfo  {journal} {International Journal of Modern
  Physics B}\ }\textbf {\bibinfo {volume} {24}},\ \bibinfo {pages} {4371}
  (\bibinfo {year} {2010})}\BibitemShut {NoStop}%
\bibitem [{\citenamefont {Carollo}\ \emph {et~al.}(2020)\citenamefont
  {Carollo}, \citenamefont {Valenti},\ and\ \citenamefont
  {Spagnolo}}]{CAROLLO20201}%
  \BibitemOpen
  \bibfield  {author} {\bibinfo {author} {\bibfnamefont {A.}~\bibnamefont
  {Carollo}}, \bibinfo {author} {\bibfnamefont {D.}~\bibnamefont {Valenti}},\
  and\ \bibinfo {author} {\bibfnamefont {B.}~\bibnamefont {Spagnolo}},\
  }\bibfield  {title} {\bibinfo {title} {Geometry of quantum phase
  transitions},\ }\href
  {https://doi.org/https://doi.org/10.1016/j.physrep.2019.11.002} {\bibfield
  {journal} {\bibinfo  {journal} {Physics Reports}\ }\textbf {\bibinfo {volume}
  {838}},\ \bibinfo {pages} {1} (\bibinfo {year} {2020})}\BibitemShut {NoStop}%
\bibitem [{\citenamefont {Arnold}\ \emph {et~al.}(2024)\citenamefont {Arnold},
  \citenamefont {Sch\"afer}, \citenamefont {Edelman},\ and\ \citenamefont
  {Bruder}}]{2024Arnold_prl}%
  \BibitemOpen
  \bibfield  {author} {\bibinfo {author} {\bibfnamefont {J.}~\bibnamefont
  {Arnold}}, \bibinfo {author} {\bibfnamefont {F.}~\bibnamefont {Sch\"afer}},
  \bibinfo {author} {\bibfnamefont {A.}~\bibnamefont {Edelman}},\ and\ \bibinfo
  {author} {\bibfnamefont {C.}~\bibnamefont {Bruder}},\ }\bibfield  {title}
  {\bibinfo {title} {{Mapping Out Phase Diagrams with Generative
  Classifiers}},\ }\href {https://doi.org/10.1103/PhysRevLett.132.207301}
  {\bibfield  {journal} {\bibinfo  {journal} {Phys. Rev. Lett.}\ }\textbf
  {\bibinfo {volume} {132}},\ \bibinfo {pages} {207301} (\bibinfo {year}
  {2024})}\BibitemShut {NoStop}%
\bibitem [{\citenamefont {Arnold}\ \emph {et~al.}(2023)\citenamefont {Arnold},
  \citenamefont {Lörch}, \citenamefont {Holtorf},\ and\ \citenamefont
  {Schäfer}}]{arnold2023machinelearningphasetransitions}%
  \BibitemOpen
  \bibfield  {author} {\bibinfo {author} {\bibfnamefont {J.}~\bibnamefont
  {Arnold}}, \bibinfo {author} {\bibfnamefont {N.}~\bibnamefont {Lörch}},
  \bibinfo {author} {\bibfnamefont {F.}~\bibnamefont {Holtorf}},\ and\ \bibinfo
  {author} {\bibfnamefont {F.}~\bibnamefont {Schäfer}},\ }\bibfield  {title}
  {\bibinfo {title} {{Machine learning phase transitions: Connections to the
  Fisher information}},\ }\href {https://arxiv.org/abs/2311.10710} {\bibfield
  {journal} {\bibinfo  {journal} {arXiv: 2311.10710}\ } (\bibinfo {year}
  {2023})}\BibitemShut {NoStop}%
\bibitem [{\citenamefont {Zanardi}\ and\ \citenamefont
  {Paunkovi\ifmmode~\acute{c}\else \'{c}\fi{}}(2006)}]{Zanardi2006_PRE}%
  \BibitemOpen
  \bibfield  {author} {\bibinfo {author} {\bibfnamefont {P.}~\bibnamefont
  {Zanardi}}\ and\ \bibinfo {author} {\bibfnamefont {N.}~\bibnamefont
  {Paunkovi\ifmmode~\acute{c}\else \'{c}\fi{}}},\ }\bibfield  {title} {\bibinfo
  {title} {{Ground state overlap and quantum phase transitions}},\ }\href
  {https://doi.org/10.1103/PhysRevE.74.031123} {\bibfield  {journal} {\bibinfo
  {journal} {Phys. Rev. E}\ }\textbf {\bibinfo {volume} {74}},\ \bibinfo
  {pages} {031123} (\bibinfo {year} {2006})}\BibitemShut {NoStop}%
\bibitem [{\citenamefont {Cozzini}\ \emph {et~al.}(2007)\citenamefont
  {Cozzini}, \citenamefont {Ionicioiu},\ and\ \citenamefont
  {Zanardi}}]{ZanardiPaolo2007}%
  \BibitemOpen
  \bibfield  {author} {\bibinfo {author} {\bibfnamefont {M.}~\bibnamefont
  {Cozzini}}, \bibinfo {author} {\bibfnamefont {R.}~\bibnamefont {Ionicioiu}},\
  and\ \bibinfo {author} {\bibfnamefont {P.}~\bibnamefont {Zanardi}},\
  }\bibfield  {title} {\bibinfo {title} {{Quantum fidelity and quantum phase
  transitions in matrix product states}},\ }\href
  {https://doi.org/10.1103/PhysRevB.76.104420} {\bibfield  {journal} {\bibinfo
  {journal} {Phys. Rev. B}\ }\textbf {\bibinfo {volume} {76}},\ \bibinfo
  {pages} {104420} (\bibinfo {year} {2007})}\BibitemShut {NoStop}%
\bibitem [{\citenamefont {Campos~Venuti}\ and\ \citenamefont
  {Zanardi}(2007)}]{Zanardi2007_prl}%
  \BibitemOpen
  \bibfield  {author} {\bibinfo {author} {\bibfnamefont {L.}~\bibnamefont
  {Campos~Venuti}}\ and\ \bibinfo {author} {\bibfnamefont {P.}~\bibnamefont
  {Zanardi}},\ }\bibfield  {title} {\bibinfo {title} {{Quantum Critical Scaling
  of the Geometric Tensors}},\ }\href
  {https://doi.org/10.1103/PhysRevLett.99.095701} {\bibfield  {journal}
  {\bibinfo  {journal} {Phys. Rev. Lett.}\ }\textbf {\bibinfo {volume} {99}},\
  \bibinfo {pages} {095701} (\bibinfo {year} {2007})}\BibitemShut {NoStop}%
\bibitem [{\citenamefont {Titum}\ \emph {et~al.}(2019)\citenamefont {Titum},
  \citenamefont {Iosue}, \citenamefont {Garrison}, \citenamefont {Gorshkov},\
  and\ \citenamefont {Gong}}]{2019PRL_Gong}%
  \BibitemOpen
  \bibfield  {author} {\bibinfo {author} {\bibfnamefont {P.}~\bibnamefont
  {Titum}}, \bibinfo {author} {\bibfnamefont {J.~T.}\ \bibnamefont {Iosue}},
  \bibinfo {author} {\bibfnamefont {J.~R.}\ \bibnamefont {Garrison}}, \bibinfo
  {author} {\bibfnamefont {A.~V.}\ \bibnamefont {Gorshkov}},\ and\ \bibinfo
  {author} {\bibfnamefont {Z.-X.}\ \bibnamefont {Gong}},\ }\bibfield  {title}
  {\bibinfo {title} {{Probing Ground-State Phase Transitions through Quench
  Dynamics}},\ }\href {https://doi.org/10.1103/PhysRevLett.123.115701}
  {\bibfield  {journal} {\bibinfo  {journal} {Phys. Rev. Lett.}\ }\textbf
  {\bibinfo {volume} {123}},\ \bibinfo {pages} {115701} (\bibinfo {year}
  {2019})}\BibitemShut {NoStop}%
\bibitem [{\citenamefont {Haldar}\ \emph {et~al.}(2021)\citenamefont {Haldar},
  \citenamefont {Mallayya}, \citenamefont {Heyl}, \citenamefont {Pollmann},
  \citenamefont {Rigol},\ and\ \citenamefont {Das}}]{2021Das_PRX}%
  \BibitemOpen
  \bibfield  {author} {\bibinfo {author} {\bibfnamefont {A.}~\bibnamefont
  {Haldar}}, \bibinfo {author} {\bibfnamefont {K.}~\bibnamefont {Mallayya}},
  \bibinfo {author} {\bibfnamefont {M.}~\bibnamefont {Heyl}}, \bibinfo {author}
  {\bibfnamefont {F.}~\bibnamefont {Pollmann}}, \bibinfo {author}
  {\bibfnamefont {M.}~\bibnamefont {Rigol}},\ and\ \bibinfo {author}
  {\bibfnamefont {A.}~\bibnamefont {Das}},\ }\bibfield  {title} {\bibinfo
  {title} {{Signatures of Quantum Phase Transitions after Quenches in Quantum
  Chaotic One-Dimensional Systems}},\ }\href
  {https://doi.org/10.1103/PhysRevX.11.031062} {\bibfield  {journal} {\bibinfo
  {journal} {Phys. Rev. X}\ }\textbf {\bibinfo {volume} {11}},\ \bibinfo
  {pages} {031062} (\bibinfo {year} {2021})}\BibitemShut {NoStop}%
\bibitem [{\citenamefont {Heyl}\ \emph {et~al.}(2018)\citenamefont {Heyl},
  \citenamefont {Pollmann},\ and\ \citenamefont {D\'ora}}]{2018Dora_PRL}%
  \BibitemOpen
  \bibfield  {author} {\bibinfo {author} {\bibfnamefont {M.}~\bibnamefont
  {Heyl}}, \bibinfo {author} {\bibfnamefont {F.}~\bibnamefont {Pollmann}},\
  and\ \bibinfo {author} {\bibfnamefont {B.}~\bibnamefont {D\'ora}},\
  }\bibfield  {title} {\bibinfo {title} {{Detecting Equilibrium and Dynamical
  Quantum Phase Transitions in Ising Chains via Out-of-Time-Ordered
  Correlators}},\ }\href {https://doi.org/10.1103/PhysRevLett.121.016801}
  {\bibfield  {journal} {\bibinfo  {journal} {Phys. Rev. Lett.}\ }\textbf
  {\bibinfo {volume} {121}},\ \bibinfo {pages} {016801} (\bibinfo {year}
  {2018})}\BibitemShut {NoStop}%
\bibitem [{\citenamefont {Da\ifmmode~\breve{g}\else \u{g}\fi{}}\ \emph
  {et~al.}(2019)\citenamefont {Da\ifmmode~\breve{g}\else \u{g}\fi{}},
  \citenamefont {Sun},\ and\ \citenamefont {Duan}}]{2019Duan_prl}%
  \BibitemOpen
  \bibfield  {author} {\bibinfo {author} {\bibfnamefont {C.~B.}\ \bibnamefont
  {Da\ifmmode~\breve{g}\else \u{g}\fi{}}}, \bibinfo {author} {\bibfnamefont
  {K.}~\bibnamefont {Sun}},\ and\ \bibinfo {author} {\bibfnamefont {L.-M.}\
  \bibnamefont {Duan}},\ }\bibfield  {title} {\bibinfo {title} {{Detection of
  Quantum Phases via Out-of-Time-Order Correlators}},\ }\href
  {https://doi.org/10.1103/PhysRevLett.123.140602} {\bibfield  {journal}
  {\bibinfo  {journal} {Phys. Rev. Lett.}\ }\textbf {\bibinfo {volume} {123}},\
  \bibinfo {pages} {140602} (\bibinfo {year} {2019})}\BibitemShut {NoStop}%
\bibitem [{\citenamefont {Da\ifmmode~\breve{g}\else \u{g}\fi{}}\ \emph
  {et~al.}(2023)\citenamefont {Da\ifmmode~\breve{g}\else \u{g}\fi{}},
  \citenamefont {Uhrich}, \citenamefont {Wang}, \citenamefont {McCulloch},\
  and\ \citenamefont {Halimeh}}]{2023PRB_Halimeh}%
  \BibitemOpen
  \bibfield  {author} {\bibinfo {author} {\bibfnamefont {C.~B.}\ \bibnamefont
  {Da\ifmmode~\breve{g}\else \u{g}\fi{}}}, \bibinfo {author} {\bibfnamefont
  {P.}~\bibnamefont {Uhrich}}, \bibinfo {author} {\bibfnamefont
  {Y.}~\bibnamefont {Wang}}, \bibinfo {author} {\bibfnamefont {I.~P.}\
  \bibnamefont {McCulloch}},\ and\ \bibinfo {author} {\bibfnamefont {J.~C.}\
  \bibnamefont {Halimeh}},\ }\bibfield  {title} {\bibinfo {title} {{Detecting
  quantum phase transitions in the quasistationary regime of Ising chains}},\
  }\href {https://doi.org/10.1103/PhysRevB.107.094432} {\bibfield  {journal}
  {\bibinfo  {journal} {Phys. Rev. B}\ }\textbf {\bibinfo {volume} {107}},\
  \bibinfo {pages} {094432} (\bibinfo {year} {2023})}\BibitemShut {NoStop}%
\bibitem [{\citenamefont {Bandyopadhyay}\ \emph {et~al.}(2023)\citenamefont
  {Bandyopadhyay}, \citenamefont {Polkovnikov},\ and\ \citenamefont
  {Dutta}}]{2023PRB_Dutta}%
  \BibitemOpen
  \bibfield  {author} {\bibinfo {author} {\bibfnamefont {S.}~\bibnamefont
  {Bandyopadhyay}}, \bibinfo {author} {\bibfnamefont {A.}~\bibnamefont
  {Polkovnikov}},\ and\ \bibinfo {author} {\bibfnamefont {A.}~\bibnamefont
  {Dutta}},\ }\bibfield  {title} {\bibinfo {title} {{Late-time critical
  behavior of local stringlike observables under quantum quenches}},\ }\href
  {https://doi.org/10.1103/PhysRevB.107.064105} {\bibfield  {journal} {\bibinfo
   {journal} {Phys. Rev. B}\ }\textbf {\bibinfo {volume} {107}},\ \bibinfo
  {pages} {064105} (\bibinfo {year} {2023})}\BibitemShut {NoStop}%
\bibitem [{\citenamefont {Lakkaraju}\ \emph {et~al.}(2024)\citenamefont
  {Lakkaraju}, \citenamefont {Haldar},\ and\ \citenamefont
  {Sen(De)}}]{Aditi2024_PRA}%
  \BibitemOpen
  \bibfield  {author} {\bibinfo {author} {\bibfnamefont {L.~G.~C.}\
  \bibnamefont {Lakkaraju}}, \bibinfo {author} {\bibfnamefont {S.~K.}\
  \bibnamefont {Haldar}},\ and\ \bibinfo {author} {\bibfnamefont
  {A.}~\bibnamefont {Sen(De)}},\ }\bibfield  {title} {\bibinfo {title}
  {{Predicting a topological quantum phase transition from dynamics via
  multisite entanglement}},\ }\href
  {https://doi.org/10.1103/PhysRevA.109.022436} {\bibfield  {journal} {\bibinfo
   {journal} {Phys. Rev. A}\ }\textbf {\bibinfo {volume} {109}},\ \bibinfo
  {pages} {022436} (\bibinfo {year} {2024})}\BibitemShut {NoStop}%
\bibitem [{\citenamefont {Paul}\ \emph {et~al.}(2024)\citenamefont {Paul},
  \citenamefont {Titum},\ and\ \citenamefont {Maghrebi}}]{paul2022hidden}%
  \BibitemOpen
  \bibfield  {author} {\bibinfo {author} {\bibfnamefont {S.}~\bibnamefont
  {Paul}}, \bibinfo {author} {\bibfnamefont {P.}~\bibnamefont {Titum}},\ and\
  \bibinfo {author} {\bibfnamefont {M.}~\bibnamefont {Maghrebi}},\ }\bibfield
  {title} {\bibinfo {title} {{Hidden quantum criticality and entanglement in
  quench dynamics}},\ }\href
  {https://doi.org/10.1103/PhysRevResearch.6.L032003} {\bibfield  {journal}
  {\bibinfo  {journal} {Phys. Rev. Res.}\ }\textbf {\bibinfo {volume} {6}},\
  \bibinfo {pages} {L032003} (\bibinfo {year} {2024})}\BibitemShut {NoStop}%
\bibitem [{\citenamefont {Nie}\ \emph {et~al.}(2020)\citenamefont {Nie},
  \citenamefont {Wei}, \citenamefont {Chen}, \citenamefont {Zhang},
  \citenamefont {Zhao}, \citenamefont {Qiu}, \citenamefont {Tian},
  \citenamefont {Ji}, \citenamefont {Xin}, \citenamefont {Lu},\ and\
  \citenamefont {Li}}]{Nie2020_prl}%
  \BibitemOpen
  \bibfield  {author} {\bibinfo {author} {\bibfnamefont {X.}~\bibnamefont
  {Nie}}, \bibinfo {author} {\bibfnamefont {B.-B.}\ \bibnamefont {Wei}},
  \bibinfo {author} {\bibfnamefont {X.}~\bibnamefont {Chen}}, \bibinfo {author}
  {\bibfnamefont {Z.}~\bibnamefont {Zhang}}, \bibinfo {author} {\bibfnamefont
  {X.}~\bibnamefont {Zhao}}, \bibinfo {author} {\bibfnamefont {C.}~\bibnamefont
  {Qiu}}, \bibinfo {author} {\bibfnamefont {Y.}~\bibnamefont {Tian}}, \bibinfo
  {author} {\bibfnamefont {Y.}~\bibnamefont {Ji}}, \bibinfo {author}
  {\bibfnamefont {T.}~\bibnamefont {Xin}}, \bibinfo {author} {\bibfnamefont
  {D.}~\bibnamefont {Lu}},\ and\ \bibinfo {author} {\bibfnamefont
  {J.}~\bibnamefont {Li}},\ }\bibfield  {title} {\bibinfo {title}
  {{Experimental Observation of Equilibrium and Dynamical Quantum Phase
  Transitions via Out-of-Time-Ordered Correlators}},\ }\href
  {https://doi.org/10.1103/PhysRevLett.124.250601} {\bibfield  {journal}
  {\bibinfo  {journal} {Phys. Rev. Lett.}\ }\textbf {\bibinfo {volume} {124}},\
  \bibinfo {pages} {250601} (\bibinfo {year} {2020})}\BibitemShut {NoStop}%
\bibitem [{\citenamefont {Da\ifmmode~\breve{g}\else \u{g}\fi{}}\ and\
  \citenamefont {Sun}(2021)}]{2021Sun_PRB}%
  \BibitemOpen
  \bibfield  {author} {\bibinfo {author} {\bibfnamefont {C.~B.}\ \bibnamefont
  {Da\ifmmode~\breve{g}\else \u{g}\fi{}}}\ and\ \bibinfo {author}
  {\bibfnamefont {K.}~\bibnamefont {Sun}},\ }\bibfield  {title} {\bibinfo
  {title} {{Dynamical crossover in the transient quench dynamics of short-range
  transverse-field Ising models}},\ }\href
  {https://doi.org/10.1103/PhysRevB.103.214402} {\bibfield  {journal} {\bibinfo
   {journal} {Phys. Rev. B}\ }\textbf {\bibinfo {volume} {103}},\ \bibinfo
  {pages} {214402} (\bibinfo {year} {2021})}\BibitemShut {NoStop}%
\bibitem [{\citenamefont {\ifmmode \check{Z}\else
  \v{Z}\fi{}unkovi\ifmmode~\check{c}\else \v{c}\fi{}}\ \emph
  {et~al.}(2018)\citenamefont {\ifmmode \check{Z}\else
  \v{Z}\fi{}unkovi\ifmmode~\check{c}\else \v{c}\fi{}}, \citenamefont {Heyl},
  \citenamefont {Knap},\ and\ \citenamefont {Silva}}]{2018Silva_prl}%
  \BibitemOpen
  \bibfield  {author} {\bibinfo {author} {\bibfnamefont {B.}~\bibnamefont
  {\ifmmode \check{Z}\else \v{Z}\fi{}unkovi\ifmmode~\check{c}\else
  \v{c}\fi{}}}, \bibinfo {author} {\bibfnamefont {M.}~\bibnamefont {Heyl}},
  \bibinfo {author} {\bibfnamefont {M.}~\bibnamefont {Knap}},\ and\ \bibinfo
  {author} {\bibfnamefont {A.}~\bibnamefont {Silva}},\ }\bibfield  {title}
  {\bibinfo {title} {{Dynamical Quantum Phase Transitions in Spin Chains with
  Long-Range Interactions: Merging Different Concepts of Nonequilibrium
  Criticality}},\ }\href {https://doi.org/10.1103/PhysRevLett.120.130601}
  {\bibfield  {journal} {\bibinfo  {journal} {Phys. Rev. Lett.}\ }\textbf
  {\bibinfo {volume} {120}},\ \bibinfo {pages} {130601} (\bibinfo {year}
  {2018})}\BibitemShut {NoStop}%
\bibitem [{\citenamefont {Jafari}\ and\ \citenamefont
  {Akbari}(2020)}]{Jafari2020_PRA}%
  \BibitemOpen
  \bibfield  {author} {\bibinfo {author} {\bibfnamefont {R.}~\bibnamefont
  {Jafari}}\ and\ \bibinfo {author} {\bibfnamefont {A.}~\bibnamefont
  {Akbari}},\ }\bibfield  {title} {\bibinfo {title} {{Dynamics of quantum
  coherence and quantum Fisher information after a sudden quench}},\ }\href
  {https://doi.org/10.1103/PhysRevA.101.062105} {\bibfield  {journal} {\bibinfo
   {journal} {Phys. Rev. A}\ }\textbf {\bibinfo {volume} {101}},\ \bibinfo
  {pages} {062105} (\bibinfo {year} {2020})}\BibitemShut {NoStop}%
\bibitem [{\citenamefont {Kobayashi}\ and\ \citenamefont
  {Motome}(2024)}]{kobayashi2024quantum}%
  \BibitemOpen
  \bibfield  {author} {\bibinfo {author} {\bibfnamefont {K.}~\bibnamefont
  {Kobayashi}}\ and\ \bibinfo {author} {\bibfnamefont {Y.}~\bibnamefont
  {Motome}},\ }\bibfield  {title} {\bibinfo {title} {Quantum reservoir probing
  of quantum phase transitions},\ }\href {https://arxiv.org/abs/2402.07097}
  {\bibfield  {journal} {\bibinfo  {journal} {arXiv: 2402.07097}\ } (\bibinfo
  {year} {2024})}\BibitemShut {NoStop}%
\bibitem [{\citenamefont {Zanardi}\ \emph {et~al.}(2008)\citenamefont
  {Zanardi}, \citenamefont {Paris},\ and\ \citenamefont
  {Campos~Venuti}}]{2008Zanardi_pra}%
  \BibitemOpen
  \bibfield  {author} {\bibinfo {author} {\bibfnamefont {P.}~\bibnamefont
  {Zanardi}}, \bibinfo {author} {\bibfnamefont {M.~G.~A.}\ \bibnamefont
  {Paris}},\ and\ \bibinfo {author} {\bibfnamefont {L.}~\bibnamefont
  {Campos~Venuti}},\ }\bibfield  {title} {\bibinfo {title} {{Quantum
  criticality as a resource for quantum estimation}},\ }\href
  {https://doi.org/10.1103/PhysRevA.78.042105} {\bibfield  {journal} {\bibinfo
  {journal} {Phys. Rev. A}\ }\textbf {\bibinfo {volume} {78}},\ \bibinfo
  {pages} {042105} (\bibinfo {year} {2008})}\BibitemShut {NoStop}%
\bibitem [{\citenamefont {Invernizzi}\ \emph {et~al.}(2008)\citenamefont
  {Invernizzi}, \citenamefont {Korbman}, \citenamefont {Campos~Venuti},\ and\
  \citenamefont {Paris}}]{Invernizzi2008PRA}%
  \BibitemOpen
  \bibfield  {author} {\bibinfo {author} {\bibfnamefont {C.}~\bibnamefont
  {Invernizzi}}, \bibinfo {author} {\bibfnamefont {M.}~\bibnamefont {Korbman}},
  \bibinfo {author} {\bibfnamefont {L.}~\bibnamefont {Campos~Venuti}},\ and\
  \bibinfo {author} {\bibfnamefont {M.~G.~A.}\ \bibnamefont {Paris}},\
  }\bibfield  {title} {\bibinfo {title} {{Optimal quantum estimation in spin
  systems at criticality}},\ }\href
  {https://doi.org/10.1103/PhysRevA.78.042106} {\bibfield  {journal} {\bibinfo
  {journal} {Phys. Rev. A}\ }\textbf {\bibinfo {volume} {78}},\ \bibinfo
  {pages} {042106} (\bibinfo {year} {2008})}\BibitemShut {NoStop}%
\bibitem [{\citenamefont {Mehboudi}\ \emph {et~al.}(2016)\citenamefont
  {Mehboudi}, \citenamefont {Correa},\ and\ \citenamefont
  {Sanpera}}]{Mehboudi2016PRA}%
  \BibitemOpen
  \bibfield  {author} {\bibinfo {author} {\bibfnamefont {M.}~\bibnamefont
  {Mehboudi}}, \bibinfo {author} {\bibfnamefont {L.~A.}\ \bibnamefont
  {Correa}},\ and\ \bibinfo {author} {\bibfnamefont {A.}~\bibnamefont
  {Sanpera}},\ }\bibfield  {title} {\bibinfo {title} {{Achieving sub-shot-noise
  sensing at finite temperatures}},\ }\href
  {https://doi.org/10.1103/PhysRevA.94.042121} {\bibfield  {journal} {\bibinfo
  {journal} {Phys. Rev. A}\ }\textbf {\bibinfo {volume} {94}},\ \bibinfo
  {pages} {042121} (\bibinfo {year} {2016})}\BibitemShut {NoStop}%
\bibitem [{\citenamefont {Rams}\ \emph {et~al.}(2018)\citenamefont {Rams},
  \citenamefont {Sierant}, \citenamefont {Dutta}, \citenamefont {Horodecki},\
  and\ \citenamefont {Zakrzewski}}]{2018prx_Marek}%
  \BibitemOpen
  \bibfield  {author} {\bibinfo {author} {\bibfnamefont {M.~M.}\ \bibnamefont
  {Rams}}, \bibinfo {author} {\bibfnamefont {P.}~\bibnamefont {Sierant}},
  \bibinfo {author} {\bibfnamefont {O.}~\bibnamefont {Dutta}}, \bibinfo
  {author} {\bibfnamefont {P.}~\bibnamefont {Horodecki}},\ and\ \bibinfo
  {author} {\bibfnamefont {J.}~\bibnamefont {Zakrzewski}},\ }\bibfield  {title}
  {\bibinfo {title} {{At the Limits of Criticality-Based Quantum Metrology:
  Apparent Super-Heisenberg Scaling Revisited}},\ }\href
  {https://doi.org/10.1103/PhysRevX.8.021022} {\bibfield  {journal} {\bibinfo
  {journal} {Phys. Rev. X}\ }\textbf {\bibinfo {volume} {8}},\ \bibinfo {pages}
  {021022} (\bibinfo {year} {2018})}\BibitemShut {NoStop}%
\bibitem [{\citenamefont {Chu}\ \emph {et~al.}(2021)\citenamefont {Chu},
  \citenamefont {Zhang}, \citenamefont {Yu},\ and\ \citenamefont
  {Cai}}]{2021Chu_prl}%
  \BibitemOpen
  \bibfield  {author} {\bibinfo {author} {\bibfnamefont {Y.}~\bibnamefont
  {Chu}}, \bibinfo {author} {\bibfnamefont {S.}~\bibnamefont {Zhang}}, \bibinfo
  {author} {\bibfnamefont {B.}~\bibnamefont {Yu}},\ and\ \bibinfo {author}
  {\bibfnamefont {J.}~\bibnamefont {Cai}},\ }\bibfield  {title} {\bibinfo
  {title} {{Dynamic Framework for Criticality-Enhanced Quantum Sensing}},\
  }\href {https://doi.org/10.1103/PhysRevLett.126.010502} {\bibfield  {journal}
  {\bibinfo  {journal} {Phys. Rev. Lett.}\ }\textbf {\bibinfo {volume} {126}},\
  \bibinfo {pages} {010502} (\bibinfo {year} {2021})}\BibitemShut {NoStop}%
\bibitem [{\citenamefont {Luo}\ \emph {et~al.}(2017)\citenamefont {Luo},
  \citenamefont {Zou}, \citenamefont {Wu}, \citenamefont {Liu}, \citenamefont
  {Han}, \citenamefont {Tey},\ and\ \citenamefont {You}}]{You2017_Science}%
  \BibitemOpen
  \bibfield  {author} {\bibinfo {author} {\bibfnamefont {X.-Y.}\ \bibnamefont
  {Luo}}, \bibinfo {author} {\bibfnamefont {Y.-Q.}\ \bibnamefont {Zou}},
  \bibinfo {author} {\bibfnamefont {L.-N.}\ \bibnamefont {Wu}}, \bibinfo
  {author} {\bibfnamefont {Q.}~\bibnamefont {Liu}}, \bibinfo {author}
  {\bibfnamefont {M.-F.}\ \bibnamefont {Han}}, \bibinfo {author} {\bibfnamefont
  {M.~K.}\ \bibnamefont {Tey}},\ and\ \bibinfo {author} {\bibfnamefont
  {L.}~\bibnamefont {You}},\ }\bibfield  {title} {\bibinfo {title}
  {{Deterministic entanglement generation from driving through quantum phase
  transitions}},\ }\href {https://doi.org/10.1126/science.aag1106} {\bibfield
  {journal} {\bibinfo  {journal} {Science}\ }\textbf {\bibinfo {volume}
  {355}},\ \bibinfo {pages} {620} (\bibinfo {year} {2017})}\BibitemShut
  {NoStop}%
\bibitem [{\citenamefont {Fr\'erot}\ and\ \citenamefont
  {Roscilde}(2018)}]{2018Tommaso_prl}%
  \BibitemOpen
  \bibfield  {author} {\bibinfo {author} {\bibfnamefont {I.}~\bibnamefont
  {Fr\'erot}}\ and\ \bibinfo {author} {\bibfnamefont {T.}~\bibnamefont
  {Roscilde}},\ }\bibfield  {title} {\bibinfo {title} {{Quantum Critical
  Metrology}},\ }\href {https://doi.org/10.1103/PhysRevLett.121.020402}
  {\bibfield  {journal} {\bibinfo  {journal} {Phys. Rev. Lett.}\ }\textbf
  {\bibinfo {volume} {121}},\ \bibinfo {pages} {020402} (\bibinfo {year}
  {2018})}\BibitemShut {NoStop}%
\bibitem [{\citenamefont {Chu}\ \emph {et~al.}(2020)\citenamefont {Chu},
  \citenamefont {Liu}, \citenamefont {Liu},\ and\ \citenamefont
  {Cai}}]{Chu2020_prl}%
  \BibitemOpen
  \bibfield  {author} {\bibinfo {author} {\bibfnamefont {Y.}~\bibnamefont
  {Chu}}, \bibinfo {author} {\bibfnamefont {Y.}~\bibnamefont {Liu}}, \bibinfo
  {author} {\bibfnamefont {H.}~\bibnamefont {Liu}},\ and\ \bibinfo {author}
  {\bibfnamefont {J.}~\bibnamefont {Cai}},\ }\bibfield  {title} {\bibinfo
  {title} {{Quantum Sensing with a Single-Qubit Pseudo-Hermitian System}},\
  }\href {https://doi.org/10.1103/PhysRevLett.124.020501} {\bibfield  {journal}
  {\bibinfo  {journal} {Phys. Rev. Lett.}\ }\textbf {\bibinfo {volume} {124}},\
  \bibinfo {pages} {020501} (\bibinfo {year} {2020})}\BibitemShut {NoStop}%
\bibitem [{\citenamefont {Ding}\ \emph {et~al.}(2022)\citenamefont {Ding},
  \citenamefont {Liu}, \citenamefont {Shi}, \citenamefont {Guo}, \citenamefont
  {M{\o}lmer},\ and\ \citenamefont {Adams}}]{ding2022enhanced}%
  \BibitemOpen
  \bibfield  {author} {\bibinfo {author} {\bibfnamefont {D.-S.}\ \bibnamefont
  {Ding}}, \bibinfo {author} {\bibfnamefont {Z.-K.}\ \bibnamefont {Liu}},
  \bibinfo {author} {\bibfnamefont {B.-S.}\ \bibnamefont {Shi}}, \bibinfo
  {author} {\bibfnamefont {G.-C.}\ \bibnamefont {Guo}}, \bibinfo {author}
  {\bibfnamefont {K.}~\bibnamefont {M{\o}lmer}},\ and\ \bibinfo {author}
  {\bibfnamefont {C.~S.}\ \bibnamefont {Adams}},\ }\bibfield  {title} {\bibinfo
  {title} {{Enhanced metrology at the critical point of a many-body Rydberg
  atomic system}},\ }\href {https://doi.org/10.1038/s41567-022-01777-8}
  {\bibfield  {journal} {\bibinfo  {journal} {Nature Physics}\ }\textbf
  {\bibinfo {volume} {18}},\ \bibinfo {pages} {1447} (\bibinfo {year}
  {2022})}\BibitemShut {NoStop}%
\bibitem [{\citenamefont {Hotter}\ \emph {et~al.}(2024)\citenamefont {Hotter},
  \citenamefont {Ritsch},\ and\ \citenamefont {Gietka}}]{2024Hotter_prl}%
  \BibitemOpen
  \bibfield  {author} {\bibinfo {author} {\bibfnamefont {C.}~\bibnamefont
  {Hotter}}, \bibinfo {author} {\bibfnamefont {H.}~\bibnamefont {Ritsch}},\
  and\ \bibinfo {author} {\bibfnamefont {K.}~\bibnamefont {Gietka}},\
  }\bibfield  {title} {\bibinfo {title} {{Combining Critical and Quantum
  Metrology}},\ }\href {https://doi.org/10.1103/PhysRevLett.132.060801}
  {\bibfield  {journal} {\bibinfo  {journal} {Phys. Rev. Lett.}\ }\textbf
  {\bibinfo {volume} {132}},\ \bibinfo {pages} {060801} (\bibinfo {year}
  {2024})}\BibitemShut {NoStop}%
\bibitem [{\citenamefont {Braun}\ \emph {et~al.}(2018)\citenamefont {Braun},
  \citenamefont {Adesso}, \citenamefont {Benatti}, \citenamefont {Floreanini},
  \citenamefont {Marzolino}, \citenamefont {Mitchell},\ and\ \citenamefont
  {Pirandola}}]{Braun2018_RMP}%
  \BibitemOpen
  \bibfield  {author} {\bibinfo {author} {\bibfnamefont {D.}~\bibnamefont
  {Braun}}, \bibinfo {author} {\bibfnamefont {G.}~\bibnamefont {Adesso}},
  \bibinfo {author} {\bibfnamefont {F.}~\bibnamefont {Benatti}}, \bibinfo
  {author} {\bibfnamefont {R.}~\bibnamefont {Floreanini}}, \bibinfo {author}
  {\bibfnamefont {U.}~\bibnamefont {Marzolino}}, \bibinfo {author}
  {\bibfnamefont {M.~W.}\ \bibnamefont {Mitchell}},\ and\ \bibinfo {author}
  {\bibfnamefont {S.}~\bibnamefont {Pirandola}},\ }\bibfield  {title} {\bibinfo
  {title} {{Quantum-enhanced measurements without entanglement}},\ }\href
  {https://doi.org/10.1103/RevModPhys.90.035006} {\bibfield  {journal}
  {\bibinfo  {journal} {Rev. Mod. Phys.}\ }\textbf {\bibinfo {volume} {90}},\
  \bibinfo {pages} {035006} (\bibinfo {year} {2018})}\BibitemShut {NoStop}%
\bibitem [{\citenamefont {Zhuang}\ \emph {et~al.}(2020)\citenamefont {Zhuang},
  \citenamefont {Huang}, \citenamefont {Ke},\ and\ \citenamefont
  {Lee}}]{Lee2020}%
  \BibitemOpen
  \bibfield  {author} {\bibinfo {author} {\bibfnamefont {M.}~\bibnamefont
  {Zhuang}}, \bibinfo {author} {\bibfnamefont {J.}~\bibnamefont {Huang}},
  \bibinfo {author} {\bibfnamefont {Y.}~\bibnamefont {Ke}},\ and\ \bibinfo
  {author} {\bibfnamefont {C.}~\bibnamefont {Lee}},\ }\bibfield  {title}
  {\bibinfo {title} {Symmetry-protected quantum adiabatic evolution in
  spontaneous symmetry-breaking transitions},\ }\href
  {https://doi.org/https://doi.org/10.1002/andp.201900471} {\bibfield
  {journal} {\bibinfo  {journal} {Annalen der Physik}\ }\textbf {\bibinfo
  {volume} {532}},\ \bibinfo {pages} {1900471} (\bibinfo {year}
  {2020})}\BibitemShut {NoStop}%
\bibitem [{\citenamefont {Chu}\ \emph {et~al.}(2023)\citenamefont {Chu},
  \citenamefont {Li},\ and\ \citenamefont {Cai}}]{2023Chu_prl}%
  \BibitemOpen
  \bibfield  {author} {\bibinfo {author} {\bibfnamefont {Y.}~\bibnamefont
  {Chu}}, \bibinfo {author} {\bibfnamefont {X.}~\bibnamefont {Li}},\ and\
  \bibinfo {author} {\bibfnamefont {J.}~\bibnamefont {Cai}},\ }\bibfield
  {title} {\bibinfo {title} {{Strong Quantum Metrological Limit from Many-Body
  Physics}},\ }\href {https://doi.org/10.1103/PhysRevLett.130.170801}
  {\bibfield  {journal} {\bibinfo  {journal} {Phys. Rev. Lett.}\ }\textbf
  {\bibinfo {volume} {130}},\ \bibinfo {pages} {170801} (\bibinfo {year}
  {2023})}\BibitemShut {NoStop}%
\bibitem [{\citenamefont {Yu}\ \emph {et~al.}(2024)\citenamefont {Yu},
  \citenamefont {Li}, \citenamefont {Chu}, \citenamefont {Mera}, \citenamefont
  {Ünal}, \citenamefont {Yang}, \citenamefont {Liu}, \citenamefont {Goldman},\
  and\ \citenamefont {Cai}}]{2024Yu_nsr}%
  \BibitemOpen
  \bibfield  {author} {\bibinfo {author} {\bibfnamefont {M.}~\bibnamefont
  {Yu}}, \bibinfo {author} {\bibfnamefont {X.}~\bibnamefont {Li}}, \bibinfo
  {author} {\bibfnamefont {Y.}~\bibnamefont {Chu}}, \bibinfo {author}
  {\bibfnamefont {B.}~\bibnamefont {Mera}}, \bibinfo {author} {\bibfnamefont
  {F.~N.}\ \bibnamefont {Ünal}}, \bibinfo {author} {\bibfnamefont
  {P.}~\bibnamefont {Yang}}, \bibinfo {author} {\bibfnamefont {Y.}~\bibnamefont
  {Liu}}, \bibinfo {author} {\bibfnamefont {N.}~\bibnamefont {Goldman}},\ and\
  \bibinfo {author} {\bibfnamefont {J.}~\bibnamefont {Cai}},\ }\bibfield
  {title} {\bibinfo {title} {{Experimental demonstration of topological bounds
  in quantum metrology}},\ }\href {https://doi.org/10.1093/nsr/nwae065}
  {\bibfield  {journal} {\bibinfo  {journal} {National Science Review}\ ,\
  \bibinfo {pages} {nwae065}} (\bibinfo {year} {2024})}\BibitemShut {NoStop}%
\bibitem [{\citenamefont {Shi}\ \emph {et~al.}(2024)\citenamefont {Shi},
  \citenamefont {Guan},\ and\ \citenamefont {Yang}}]{Guan2024_prl}%
  \BibitemOpen
  \bibfield  {author} {\bibinfo {author} {\bibfnamefont {H.-L.}\ \bibnamefont
  {Shi}}, \bibinfo {author} {\bibfnamefont {X.-W.}\ \bibnamefont {Guan}},\ and\
  \bibinfo {author} {\bibfnamefont {J.}~\bibnamefont {Yang}},\ }\bibfield
  {title} {\bibinfo {title} {{Universal Shot-Noise Limit for Quantum Metrology
  with Local Hamiltonians}},\ }\href
  {https://doi.org/10.1103/PhysRevLett.132.100803} {\bibfield  {journal}
  {\bibinfo  {journal} {Phys. Rev. Lett.}\ }\textbf {\bibinfo {volume} {132}},\
  \bibinfo {pages} {100803} (\bibinfo {year} {2024})}\BibitemShut {NoStop}%
\bibitem [{\citenamefont {Block}\ \emph {et~al.}(2024)\citenamefont {Block},
  \citenamefont {Ye}, \citenamefont {Roberts}, \citenamefont {Chern},
  \citenamefont {Wu}, \citenamefont {Wang}, \citenamefont {Pollet},
  \citenamefont {Davis}, \citenamefont {Halperin},\ and\ \citenamefont
  {Yao}}]{Block2024Scalable}%
  \BibitemOpen
  \bibfield  {author} {\bibinfo {author} {\bibfnamefont {M.}~\bibnamefont
  {Block}}, \bibinfo {author} {\bibfnamefont {B.}~\bibnamefont {Ye}}, \bibinfo
  {author} {\bibfnamefont {B.}~\bibnamefont {Roberts}}, \bibinfo {author}
  {\bibfnamefont {S.}~\bibnamefont {Chern}}, \bibinfo {author} {\bibfnamefont
  {W.}~\bibnamefont {Wu}}, \bibinfo {author} {\bibfnamefont {Z.}~\bibnamefont
  {Wang}}, \bibinfo {author} {\bibfnamefont {L.}~\bibnamefont {Pollet}},
  \bibinfo {author} {\bibfnamefont {E.~J.}\ \bibnamefont {Davis}}, \bibinfo
  {author} {\bibfnamefont {B.~I.}\ \bibnamefont {Halperin}},\ and\ \bibinfo
  {author} {\bibfnamefont {N.~Y.}\ \bibnamefont {Yao}},\ }\bibfield  {title}
  {\bibinfo {title} {Scalable spin squeezing from finite-temperature easy-plane
  magnetism},\ }\href {https://doi.org/10.1038/s41567-024-02562-5} {\bibfield
  {journal} {\bibinfo  {journal} {Nature Physics}\ }\textbf {\bibinfo {volume}
  {20}},\ \bibinfo {pages} {1575} (\bibinfo {year} {2024})}\BibitemShut
  {NoStop}%
\bibitem [{\citenamefont {Mihailescu}\ \emph
  {et~al.}(2024{\natexlab{a}})\citenamefont {Mihailescu}, \citenamefont
  {Bayat}, \citenamefont {Campbell},\ and\ \citenamefont
  {Mitchell}}]{Mihailescu_2024Multiparameter}%
  \BibitemOpen
  \bibfield  {author} {\bibinfo {author} {\bibfnamefont {G.}~\bibnamefont
  {Mihailescu}}, \bibinfo {author} {\bibfnamefont {A.}~\bibnamefont {Bayat}},
  \bibinfo {author} {\bibfnamefont {S.}~\bibnamefont {Campbell}},\ and\
  \bibinfo {author} {\bibfnamefont {A.~K.}\ \bibnamefont {Mitchell}},\
  }\bibfield  {title} {\bibinfo {title} {{Multiparameter critical quantum
  metrology with impurity probes}},\ }\href
  {https://doi.org/10.1088/2058-9565/ad438d} {\bibfield  {journal} {\bibinfo
  {journal} {Quantum Science and Technology}\ }\textbf {\bibinfo {volume}
  {9}},\ \bibinfo {pages} {035033} (\bibinfo {year}
  {2024}{\natexlab{a}})}\BibitemShut {NoStop}%
\bibitem [{\citenamefont {Mihailescu}\ \emph
  {et~al.}(2024{\natexlab{b}})\citenamefont {Mihailescu}, \citenamefont
  {Campbell},\ and\ \citenamefont
  {Gietka}}]{mihailescu2024uncertainquantumcriticalmetrology}%
  \BibitemOpen
  \bibfield  {author} {\bibinfo {author} {\bibfnamefont {G.}~\bibnamefont
  {Mihailescu}}, \bibinfo {author} {\bibfnamefont {S.}~\bibnamefont
  {Campbell}},\ and\ \bibinfo {author} {\bibfnamefont {K.}~\bibnamefont
  {Gietka}},\ }\bibfield  {title} {\bibinfo {title} {{Uncertain Quantum
  Critical Metrology: From Single to Multi Parameter Sensing}},\ }\href
  {https://arxiv.org/abs/2407.19917} {\bibfield  {journal} {\bibinfo  {journal}
  {arXiv: 2407.19917}\ } (\bibinfo {year} {2024}{\natexlab{b}})}\BibitemShut
  {NoStop}%
\bibitem [{\citenamefont {Adani}\ \emph {et~al.}(2024)\citenamefont {Adani},
  \citenamefont {Cavazzoni}, \citenamefont {Teklu}, \citenamefont {Bordone},\
  and\ \citenamefont {Paris}}]{adani2024criticalmetrologyminimallyaccessible}%
  \BibitemOpen
  \bibfield  {author} {\bibinfo {author} {\bibfnamefont {M.}~\bibnamefont
  {Adani}}, \bibinfo {author} {\bibfnamefont {S.}~\bibnamefont {Cavazzoni}},
  \bibinfo {author} {\bibfnamefont {B.}~\bibnamefont {Teklu}}, \bibinfo
  {author} {\bibfnamefont {P.}~\bibnamefont {Bordone}},\ and\ \bibinfo {author}
  {\bibfnamefont {M.~G.~A.}\ \bibnamefont {Paris}},\ }\bibfield  {title}
  {\bibinfo {title} {Critical metrology of minimally accessible anisotropic
  spin chains},\ }\href {https://arxiv.org/abs/2405.20296} {\bibfield
  {journal} {\bibinfo  {journal} {arXiv: 2405.20296}\ } (\bibinfo {year}
  {2024})}\BibitemShut {NoStop}%
\bibitem [{\citenamefont {Braunstein}\ and\ \citenamefont
  {Caves}(1994)}]{Caves1994_prl}%
  \BibitemOpen
  \bibfield  {author} {\bibinfo {author} {\bibfnamefont {S.~L.}\ \bibnamefont
  {Braunstein}}\ and\ \bibinfo {author} {\bibfnamefont {C.~M.}\ \bibnamefont
  {Caves}},\ }\bibfield  {title} {\bibinfo {title} {{Statistical distance and
  the geometry of quantum states}},\ }\href
  {https://doi.org/10.1103/PhysRevLett.72.3439} {\bibfield  {journal} {\bibinfo
   {journal} {Phys. Rev. Lett.}\ }\textbf {\bibinfo {volume} {72}},\ \bibinfo
  {pages} {3439} (\bibinfo {year} {1994})}\BibitemShut {NoStop}%
\bibitem [{\citenamefont {Pezz\`e}\ \emph {et~al.}(2018)\citenamefont
  {Pezz\`e}, \citenamefont {Smerzi}, \citenamefont {Oberthaler}, \citenamefont
  {Schmied},\ and\ \citenamefont {Treutlein}}]{2018RMP_Pezze}%
  \BibitemOpen
  \bibfield  {author} {\bibinfo {author} {\bibfnamefont {L.}~\bibnamefont
  {Pezz\`e}}, \bibinfo {author} {\bibfnamefont {A.}~\bibnamefont {Smerzi}},
  \bibinfo {author} {\bibfnamefont {M.~K.}\ \bibnamefont {Oberthaler}},
  \bibinfo {author} {\bibfnamefont {R.}~\bibnamefont {Schmied}},\ and\ \bibinfo
  {author} {\bibfnamefont {P.}~\bibnamefont {Treutlein}},\ }\bibfield  {title}
  {\bibinfo {title} {{Quantum metrology with nonclassical states of atomic
  ensembles}},\ }\href {https://doi.org/10.1103/RevModPhys.90.035005}
  {\bibfield  {journal} {\bibinfo  {journal} {Rev. Mod. Phys.}\ }\textbf
  {\bibinfo {volume} {90}},\ \bibinfo {pages} {035005} (\bibinfo {year}
  {2018})}\BibitemShut {NoStop}%
\bibitem [{\citenamefont {Liu}\ \emph {et~al.}(2019)\citenamefont {Liu},
  \citenamefont {Yuan}, \citenamefont {Lu},\ and\ \citenamefont
  {Wang}}]{Liu_2020}%
  \BibitemOpen
  \bibfield  {author} {\bibinfo {author} {\bibfnamefont {J.}~\bibnamefont
  {Liu}}, \bibinfo {author} {\bibfnamefont {H.}~\bibnamefont {Yuan}}, \bibinfo
  {author} {\bibfnamefont {X.-M.}\ \bibnamefont {Lu}},\ and\ \bibinfo {author}
  {\bibfnamefont {X.}~\bibnamefont {Wang}},\ }\bibfield  {title} {\bibinfo
  {title} {{Quantum Fisher information matrix and multiparameter estimation}},\
  }\href {https://doi.org/10.1088/1751-8121/ab5d4d} {\bibfield  {journal}
  {\bibinfo  {journal} {Journal of Physics A: Mathematical and Theoretical}\
  }\textbf {\bibinfo {volume} {53}},\ \bibinfo {pages} {023001} (\bibinfo
  {year} {2019})}\BibitemShut {NoStop}%
\bibitem [{\citenamefont {Giovannetti}\ \emph {et~al.}(2006)\citenamefont
  {Giovannetti}, \citenamefont {Lloyd},\ and\ \citenamefont
  {Maccone}}]{2006prl_quantum_metrology}%
  \BibitemOpen
  \bibfield  {author} {\bibinfo {author} {\bibfnamefont {V.}~\bibnamefont
  {Giovannetti}}, \bibinfo {author} {\bibfnamefont {S.}~\bibnamefont {Lloyd}},\
  and\ \bibinfo {author} {\bibfnamefont {L.}~\bibnamefont {Maccone}},\
  }\bibfield  {title} {\bibinfo {title} {{Quantum Metrology}},\ }\href
  {https://doi.org/10.1103/PhysRevLett.96.010401} {\bibfield  {journal}
  {\bibinfo  {journal} {Phys. Rev. Lett.}\ }\textbf {\bibinfo {volume} {96}},\
  \bibinfo {pages} {010401} (\bibinfo {year} {2006})}\BibitemShut {NoStop}%
\bibitem [{\citenamefont {Giovannetti}\ \emph {et~al.}(2011)\citenamefont
  {Giovannetti}, \citenamefont {Lloyd},\ and\ \citenamefont
  {Maccone}}]{giovannetti2011advances}%
  \BibitemOpen
  \bibfield  {author} {\bibinfo {author} {\bibfnamefont {V.}~\bibnamefont
  {Giovannetti}}, \bibinfo {author} {\bibfnamefont {S.}~\bibnamefont {Lloyd}},\
  and\ \bibinfo {author} {\bibfnamefont {L.}~\bibnamefont {Maccone}},\
  }\bibfield  {title} {\bibinfo {title} {{Advances in quantum metrology}},\
  }\href {https://doi.org/10.1038/nphoton.2011.35} {\bibfield  {journal}
  {\bibinfo  {journal} {Nature photonics}\ }\textbf {\bibinfo {volume} {5}},\
  \bibinfo {pages} {222} (\bibinfo {year} {2011})}\BibitemShut {NoStop}%
\bibitem [{\citenamefont {Wineland}\ \emph {et~al.}(1994)\citenamefont
  {Wineland}, \citenamefont {Bollinger}, \citenamefont {Itano},\ and\
  \citenamefont {Heinzen}}]{Wineland1994PRA}%
  \BibitemOpen
  \bibfield  {author} {\bibinfo {author} {\bibfnamefont {D.~J.}\ \bibnamefont
  {Wineland}}, \bibinfo {author} {\bibfnamefont {J.~J.}\ \bibnamefont
  {Bollinger}}, \bibinfo {author} {\bibfnamefont {W.~M.}\ \bibnamefont
  {Itano}},\ and\ \bibinfo {author} {\bibfnamefont {D.~J.}\ \bibnamefont
  {Heinzen}},\ }\bibfield  {title} {\bibinfo {title} {{Squeezed atomic states
  and projection noise in spectroscopy}},\ }\href
  {https://doi.org/10.1103/PhysRevA.50.67} {\bibfield  {journal} {\bibinfo
  {journal} {Phys. Rev. A}\ }\textbf {\bibinfo {volume} {50}},\ \bibinfo
  {pages} {67} (\bibinfo {year} {1994})}\BibitemShut {NoStop}%
\bibitem [{\citenamefont {Hauke}\ \emph {et~al.}(2016)\citenamefont {Hauke},
  \citenamefont {Heyl}, \citenamefont {Tagliacozzo},\ and\ \citenamefont
  {Zoller}}]{Zoller2016measuring}%
  \BibitemOpen
  \bibfield  {author} {\bibinfo {author} {\bibfnamefont {P.}~\bibnamefont
  {Hauke}}, \bibinfo {author} {\bibfnamefont {M.}~\bibnamefont {Heyl}},
  \bibinfo {author} {\bibfnamefont {L.}~\bibnamefont {Tagliacozzo}},\ and\
  \bibinfo {author} {\bibfnamefont {P.}~\bibnamefont {Zoller}},\ }\bibfield
  {title} {\bibinfo {title} {{Measuring multipartite entanglement through
  dynamic susceptibilities}},\ }\href {https://doi.org/10.1038/nphys3700}
  {\bibfield  {journal} {\bibinfo  {journal} {Nature Physics}\ }\textbf
  {\bibinfo {volume} {12}},\ \bibinfo {pages} {778} (\bibinfo {year}
  {2016})}\BibitemShut {NoStop}%
\bibitem [{\citenamefont {Niezgoda}\ and\ \citenamefont
  {Chwede\ifmmode~\acute{n}\else \'{n}\fi{}czuk}(2021)}]{Niezgoda2021}%
  \BibitemOpen
  \bibfield  {author} {\bibinfo {author} {\bibfnamefont {A.}~\bibnamefont
  {Niezgoda}}\ and\ \bibinfo {author} {\bibfnamefont {J.}~\bibnamefont
  {Chwede\ifmmode~\acute{n}\else \'{n}\fi{}czuk}},\ }\bibfield  {title}
  {\bibinfo {title} {{Many-Body Nonlocality as a Resource for Quantum-Enhanced
  Metrology}},\ }\href {https://doi.org/10.1103/PhysRevLett.126.210506}
  {\bibfield  {journal} {\bibinfo  {journal} {Phys. Rev. Lett.}\ }\textbf
  {\bibinfo {volume} {126}},\ \bibinfo {pages} {210506} (\bibinfo {year}
  {2021})}\BibitemShut {NoStop}%
\bibitem [{\citenamefont {Chu}\ \emph {et~al.}(2024)\citenamefont {Chu},
  \citenamefont {Li},\ and\ \citenamefont {Cai}}]{Chu2024}%
  \BibitemOpen
  \bibfield  {author} {\bibinfo {author} {\bibfnamefont {Y.}~\bibnamefont
  {Chu}}, \bibinfo {author} {\bibfnamefont {X.}~\bibnamefont {Li}},\ and\
  \bibinfo {author} {\bibfnamefont {J.}~\bibnamefont {Cai}},\ }\bibfield
  {title} {\bibinfo {title} {Quantum delocalization on correlation landscape:
  The key to exponentially fast multipartite entanglement generation},\ }\href
  {https://arxiv.org/abs/2404.10973} {\bibfield  {journal} {\bibinfo  {journal}
  {arXiv: 2404.10973}\ } (\bibinfo {year} {2024})}\BibitemShut {NoStop}%
\bibitem [{\citenamefont {Amico}\ \emph {et~al.}(2008)\citenamefont {Amico},
  \citenamefont {Fazio}, \citenamefont {Osterloh},\ and\ \citenamefont
  {Vedral}}]{2008Amico_RMP_Entanglement}%
  \BibitemOpen
  \bibfield  {author} {\bibinfo {author} {\bibfnamefont {L.}~\bibnamefont
  {Amico}}, \bibinfo {author} {\bibfnamefont {R.}~\bibnamefont {Fazio}},
  \bibinfo {author} {\bibfnamefont {A.}~\bibnamefont {Osterloh}},\ and\
  \bibinfo {author} {\bibfnamefont {V.}~\bibnamefont {Vedral}},\ }\bibfield
  {title} {\bibinfo {title} {Entanglement in many-body systems},\ }\href
  {https://doi.org/10.1103/RevModPhys.80.517} {\bibfield  {journal} {\bibinfo
  {journal} {Rev. Mod. Phys.}\ }\textbf {\bibinfo {volume} {80}},\ \bibinfo
  {pages} {517} (\bibinfo {year} {2008})}\BibitemShut {NoStop}%
\bibitem [{\citenamefont {Nahum}\ \emph {et~al.}(2017)\citenamefont {Nahum},
  \citenamefont {Ruhman}, \citenamefont {Vijay},\ and\ \citenamefont
  {Haah}}]{2017Adam_EntanglementGrowth}%
  \BibitemOpen
  \bibfield  {author} {\bibinfo {author} {\bibfnamefont {A.}~\bibnamefont
  {Nahum}}, \bibinfo {author} {\bibfnamefont {J.}~\bibnamefont {Ruhman}},
  \bibinfo {author} {\bibfnamefont {S.}~\bibnamefont {Vijay}},\ and\ \bibinfo
  {author} {\bibfnamefont {J.}~\bibnamefont {Haah}},\ }\bibfield  {title}
  {\bibinfo {title} {{Quantum Entanglement Growth under Random Unitary
  Dynamics}},\ }\href {https://doi.org/10.1103/PhysRevX.7.031016} {\bibfield
  {journal} {\bibinfo  {journal} {Phys. Rev. X}\ }\textbf {\bibinfo {volume}
  {7}},\ \bibinfo {pages} {031016} (\bibinfo {year} {2017})}\BibitemShut
  {NoStop}%
\bibitem [{\citenamefont {Baykusheva}\ \emph {et~al.}(2023)\citenamefont
  {Baykusheva}, \citenamefont {Kalthoff}, \citenamefont {Hofmann},
  \citenamefont {Claassen}, \citenamefont {Kennes}, \citenamefont {Sentef},\
  and\ \citenamefont {Mitrano}}]{Baykusheva2023}%
  \BibitemOpen
  \bibfield  {author} {\bibinfo {author} {\bibfnamefont {D.~R.}\ \bibnamefont
  {Baykusheva}}, \bibinfo {author} {\bibfnamefont {M.~H.}\ \bibnamefont
  {Kalthoff}}, \bibinfo {author} {\bibfnamefont {D.}~\bibnamefont {Hofmann}},
  \bibinfo {author} {\bibfnamefont {M.}~\bibnamefont {Claassen}}, \bibinfo
  {author} {\bibfnamefont {D.~M.}\ \bibnamefont {Kennes}}, \bibinfo {author}
  {\bibfnamefont {M.~A.}\ \bibnamefont {Sentef}},\ and\ \bibinfo {author}
  {\bibfnamefont {M.}~\bibnamefont {Mitrano}},\ }\bibfield  {title} {\bibinfo
  {title} {{Witnessing Nonequilibrium Entanglement Dynamics in a Strongly
  Correlated Fermionic Chain}},\ }\href
  {https://doi.org/10.1103/PhysRevLett.130.106902} {\bibfield  {journal}
  {\bibinfo  {journal} {Phys. Rev. Lett.}\ }\textbf {\bibinfo {volume} {130}},\
  \bibinfo {pages} {106902} (\bibinfo {year} {2023})}\BibitemShut {NoStop}%
\bibitem [{\citenamefont {G\"arttner}\ \emph {et~al.}(2018)\citenamefont
  {G\"arttner}, \citenamefont {Hauke},\ and\ \citenamefont
  {Rey}}]{Garttner2018}%
  \BibitemOpen
  \bibfield  {author} {\bibinfo {author} {\bibfnamefont {M.}~\bibnamefont
  {G\"arttner}}, \bibinfo {author} {\bibfnamefont {P.}~\bibnamefont {Hauke}},\
  and\ \bibinfo {author} {\bibfnamefont {A.~M.}\ \bibnamefont {Rey}},\
  }\bibfield  {title} {\bibinfo {title} {{Relating Out-of-Time-Order
  Correlations to Entanglement via Multiple-Quantum Coherences}},\ }\href
  {https://doi.org/10.1103/PhysRevLett.120.040402} {\bibfield  {journal}
  {\bibinfo  {journal} {Phys. Rev. Lett.}\ }\textbf {\bibinfo {volume} {120}},\
  \bibinfo {pages} {040402} (\bibinfo {year} {2018})}\BibitemShut {NoStop}%
\bibitem [{\citenamefont {Lewis-Swan}\ \emph {et~al.}(2020)\citenamefont
  {Lewis-Swan}, \citenamefont {Muleady},\ and\ \citenamefont
  {Rey}}]{LewisSwan2020}%
  \BibitemOpen
  \bibfield  {author} {\bibinfo {author} {\bibfnamefont {R.~J.}\ \bibnamefont
  {Lewis-Swan}}, \bibinfo {author} {\bibfnamefont {S.~R.}\ \bibnamefont
  {Muleady}},\ and\ \bibinfo {author} {\bibfnamefont {A.~M.}\ \bibnamefont
  {Rey}},\ }\bibfield  {title} {\bibinfo {title} {{Detecting Out-of-Time-Order
  Correlations via Quasiadiabatic Echoes as a Tool to Reveal Quantum Coherence
  in Equilibrium Quantum Phase Transitions}},\ }\href
  {https://doi.org/10.1103/PhysRevLett.125.240605} {\bibfield  {journal}
  {\bibinfo  {journal} {Phys. Rev. Lett.}\ }\textbf {\bibinfo {volume} {125}},\
  \bibinfo {pages} {240605} (\bibinfo {year} {2020})}\BibitemShut {NoStop}%
\bibitem [{\citenamefont {Karrasch}\ and\ \citenamefont
  {Schuricht}(2013)}]{2013Schuricht_prb}%
  \BibitemOpen
  \bibfield  {author} {\bibinfo {author} {\bibfnamefont {C.}~\bibnamefont
  {Karrasch}}\ and\ \bibinfo {author} {\bibfnamefont {D.}~\bibnamefont
  {Schuricht}},\ }\bibfield  {title} {\bibinfo {title} {{Dynamical phase
  transitions after quenches in nonintegrable models}},\ }\href
  {https://doi.org/10.1103/PhysRevB.87.195104} {\bibfield  {journal} {\bibinfo
  {journal} {Phys. Rev. B}\ }\textbf {\bibinfo {volume} {87}},\ \bibinfo
  {pages} {195104} (\bibinfo {year} {2013})}\BibitemShut {NoStop}%
\bibitem [{\citenamefont {Alba}\ and\ \citenamefont
  {Fagotti}(2017)}]{2017Alba_prl}%
  \BibitemOpen
  \bibfield  {author} {\bibinfo {author} {\bibfnamefont {V.}~\bibnamefont
  {Alba}}\ and\ \bibinfo {author} {\bibfnamefont {M.}~\bibnamefont {Fagotti}},\
  }\bibfield  {title} {\bibinfo {title} {{Prethermalization at Low Temperature:
  The Scent of Long-Range Order}},\ }\href
  {https://doi.org/10.1103/PhysRevLett.119.010601} {\bibfield  {journal}
  {\bibinfo  {journal} {Phys. Rev. Lett.}\ }\textbf {\bibinfo {volume} {119}},\
  \bibinfo {pages} {010601} (\bibinfo {year} {2017})}\BibitemShut {NoStop}%
\bibitem [{\citenamefont {Beccaria}\ \emph {et~al.}(2006)\citenamefont
  {Beccaria}, \citenamefont {Campostrini},\ and\ \citenamefont
  {Feo}}]{Beccaria2006_prb}%
  \BibitemOpen
  \bibfield  {author} {\bibinfo {author} {\bibfnamefont {M.}~\bibnamefont
  {Beccaria}}, \bibinfo {author} {\bibfnamefont {M.}~\bibnamefont
  {Campostrini}},\ and\ \bibinfo {author} {\bibfnamefont {A.}~\bibnamefont
  {Feo}},\ }\bibfield  {title} {\bibinfo {title} {{Density-matrix
  renormalization-group study of the disorder line in the quantum axial
  next-nearest-neighbor Ising model}},\ }\href
  {https://doi.org/10.1103/PhysRevB.73.052402} {\bibfield  {journal} {\bibinfo
  {journal} {Phys. Rev. B}\ }\textbf {\bibinfo {volume} {73}},\ \bibinfo
  {pages} {052402} (\bibinfo {year} {2006})}\BibitemShut {NoStop}%
\bibitem [{\citenamefont {Weinberg}\ and\ \citenamefont
  {Bukov}(2017)}]{quspin1}%
  \BibitemOpen
  \bibfield  {author} {\bibinfo {author} {\bibfnamefont {P.}~\bibnamefont
  {Weinberg}}\ and\ \bibinfo {author} {\bibfnamefont {M.}~\bibnamefont
  {Bukov}},\ }\bibfield  {title} {\bibinfo {title} {{QuSpin: a Python package
  for dynamics and exact diagonalisation of quantum many body systems part I:
  spin chains}},\ }\href {https://doi.org/10.21468/SciPostPhys.2.1.003}
  {\bibfield  {journal} {\bibinfo  {journal} {SciPost Phys.}\ }\textbf
  {\bibinfo {volume} {2}},\ \bibinfo {pages} {003} (\bibinfo {year}
  {2017})}\BibitemShut {NoStop}%
\bibitem [{\citenamefont {Weinberg}\ and\ \citenamefont
  {Bukov}(2019)}]{quspin2}%
  \BibitemOpen
  \bibfield  {author} {\bibinfo {author} {\bibfnamefont {P.}~\bibnamefont
  {Weinberg}}\ and\ \bibinfo {author} {\bibfnamefont {M.}~\bibnamefont
  {Bukov}},\ }\bibfield  {title} {\bibinfo {title} {{QuSpin: a Python package
  for dynamics and exact diagonalisation of quantum many body systems. Part II:
  bosons, fermions and higher spins}},\ }\href
  {https://doi.org/10.21468/SciPostPhys.7.2.020} {\bibfield  {journal}
  {\bibinfo  {journal} {SciPost Phys.}\ }\textbf {\bibinfo {volume} {7}},\
  \bibinfo {pages} {020} (\bibinfo {year} {2019})}\BibitemShut {NoStop}%
\bibitem [{sup()}]{supplement}%
  \BibitemOpen
  \href@noop {} {\bibinfo  {journal} {See Supplementary Information for
  additional details}\ }\BibitemShut {NoStop}%
\bibitem [{\citenamefont {Davis}\ \emph {et~al.}(2016)\citenamefont {Davis},
  \citenamefont {Bentsen},\ and\ \citenamefont
  {Schleier-Smith}}]{2016PRLDavis}%
  \BibitemOpen
\bibfield  {journal} {  }\bibfield  {author} {\bibinfo {author} {\bibfnamefont
  {E.}~\bibnamefont {Davis}}, \bibinfo {author} {\bibfnamefont
  {G.}~\bibnamefont {Bentsen}},\ and\ \bibinfo {author} {\bibfnamefont
  {M.}~\bibnamefont {Schleier-Smith}},\ }\bibfield  {title} {\bibinfo {title}
  {{Approaching the Heisenberg Limit without Single-Particle Detection}},\
  }\href {https://doi.org/10.1103/PhysRevLett.116.053601} {\bibfield  {journal}
  {\bibinfo  {journal} {Phys. Rev. Lett.}\ }\textbf {\bibinfo {volume} {116}},\
  \bibinfo {pages} {053601} (\bibinfo {year} {2016})}\BibitemShut {NoStop}%
\bibitem [{\citenamefont {Fr\"owis}\ \emph {et~al.}(2016)\citenamefont
  {Fr\"owis}, \citenamefont {Sekatski},\ and\ \citenamefont
  {D\"ur}}]{2016PRL_Dur}%
  \BibitemOpen
  \bibfield  {author} {\bibinfo {author} {\bibfnamefont {F.}~\bibnamefont
  {Fr\"owis}}, \bibinfo {author} {\bibfnamefont {P.}~\bibnamefont {Sekatski}},\
  and\ \bibinfo {author} {\bibfnamefont {W.}~\bibnamefont {D\"ur}},\ }\bibfield
   {title} {\bibinfo {title} {{Detecting Large Quantum Fisher Information with
  Finite Measurement Precision}},\ }\href
  {https://doi.org/10.1103/PhysRevLett.116.090801} {\bibfield  {journal}
  {\bibinfo  {journal} {Phys. Rev. Lett.}\ }\textbf {\bibinfo {volume} {116}},\
  \bibinfo {pages} {090801} (\bibinfo {year} {2016})}\BibitemShut {NoStop}%
\bibitem [{\citenamefont {Fisher}(1925)}]{fisher1925theory}%
  \BibitemOpen
  \bibfield  {author} {\bibinfo {author} {\bibfnamefont {R.~A.}\ \bibnamefont
  {Fisher}},\ }\bibfield  {title} {\bibinfo {title} {{Theory of statistical
  estimation}},\ }in\ \href@noop {} {\emph {\bibinfo {booktitle} {Mathematical
  proceedings of the Cambridge philosophical society}}},\ Vol.~\bibinfo
  {volume} {22}\ (\bibinfo {organization} {Cambridge University Press},\
  \bibinfo {year} {1925})\ pp.\ \bibinfo {pages} {700--725}\BibitemShut
  {NoStop}%
\bibitem [{\citenamefont {Braunstein}(1992)}]{Braunstein_1992}%
  \BibitemOpen
  \bibfield  {author} {\bibinfo {author} {\bibfnamefont {S.~L.}\ \bibnamefont
  {Braunstein}},\ }\bibfield  {title} {\bibinfo {title} {{How large a sample is
  needed for the maximum likelihood estimator to be approximately Gaussian?}},\
  }\href {https://doi.org/10.1088/0305-4470/25/13/027} {\bibfield  {journal}
  {\bibinfo  {journal} {Journal of Physics A: Mathematical and General}\
  }\textbf {\bibinfo {volume} {25}},\ \bibinfo {pages} {3813} (\bibinfo {year}
  {1992})}\BibitemShut {NoStop}%
\bibitem [{\citenamefont {Bloch}\ \emph {et~al.}(2008)\citenamefont {Bloch},
  \citenamefont {Dalibard},\ and\ \citenamefont {Zwerger}}]{Bloch2008_RMP}%
  \BibitemOpen
  \bibfield  {author} {\bibinfo {author} {\bibfnamefont {I.}~\bibnamefont
  {Bloch}}, \bibinfo {author} {\bibfnamefont {J.}~\bibnamefont {Dalibard}},\
  and\ \bibinfo {author} {\bibfnamefont {W.}~\bibnamefont {Zwerger}},\
  }\bibfield  {title} {\bibinfo {title} {{Many-body physics with ultracold
  gases}},\ }\href {https://doi.org/10.1103/RevModPhys.80.885} {\bibfield
  {journal} {\bibinfo  {journal} {Rev. Mod. Phys.}\ }\textbf {\bibinfo {volume}
  {80}},\ \bibinfo {pages} {885} (\bibinfo {year} {2008})}\BibitemShut
  {NoStop}%
\bibitem [{\citenamefont {Bloch}\ \emph {et~al.}(2012)\citenamefont {Bloch},
  \citenamefont {Dalibard},\ and\ \citenamefont
  {Nascimbene}}]{bloch2012quantum}%
  \BibitemOpen
  \bibfield  {author} {\bibinfo {author} {\bibfnamefont {I.}~\bibnamefont
  {Bloch}}, \bibinfo {author} {\bibfnamefont {J.}~\bibnamefont {Dalibard}},\
  and\ \bibinfo {author} {\bibfnamefont {S.}~\bibnamefont {Nascimbene}},\
  }\bibfield  {title} {\bibinfo {title} {{Quantum simulations with ultracold
  quantum gases}},\ }\href {https://doi.org/10.1038/nphys2259} {\bibfield
  {journal} {\bibinfo  {journal} {Nature Physics}\ }\textbf {\bibinfo {volume}
  {8}},\ \bibinfo {pages} {267} (\bibinfo {year} {2012})}\BibitemShut {NoStop}%
\bibitem [{\citenamefont {Gross}\ and\ \citenamefont
  {Bloch}(2017)}]{2017Bloch_ultracoldatoms}%
  \BibitemOpen
  \bibfield  {author} {\bibinfo {author} {\bibfnamefont {C.}~\bibnamefont
  {Gross}}\ and\ \bibinfo {author} {\bibfnamefont {I.}~\bibnamefont {Bloch}},\
  }\bibfield  {title} {\bibinfo {title} {Quantum simulations with ultracold
  atoms in optical lattices},\ }\href {https://doi.org/10.1126/science.aal3837}
  {\bibfield  {journal} {\bibinfo  {journal} {Science}\ }\textbf {\bibinfo
  {volume} {357}},\ \bibinfo {pages} {995} (\bibinfo {year}
  {2017})}\BibitemShut {NoStop}%
\bibitem [{\citenamefont {Sch{\"a}fer}\ \emph {et~al.}(2020)\citenamefont
  {Sch{\"a}fer}, \citenamefont {Fukuhara}, \citenamefont {Sugawa},
  \citenamefont {Takasu},\ and\ \citenamefont
  {Takahashi}}]{schafer2020tools_ultracoldatoms}%
  \BibitemOpen
  \bibfield  {author} {\bibinfo {author} {\bibfnamefont {F.}~\bibnamefont
  {Sch{\"a}fer}}, \bibinfo {author} {\bibfnamefont {T.}~\bibnamefont
  {Fukuhara}}, \bibinfo {author} {\bibfnamefont {S.}~\bibnamefont {Sugawa}},
  \bibinfo {author} {\bibfnamefont {Y.}~\bibnamefont {Takasu}},\ and\ \bibinfo
  {author} {\bibfnamefont {Y.}~\bibnamefont {Takahashi}},\ }\bibfield  {title}
  {\bibinfo {title} {Tools for quantum simulation with ultracold atoms in
  optical lattices},\ }\href {https://doi.org/10.1038/s42254-020-0195-3}
  {\bibfield  {journal} {\bibinfo  {journal} {Nature Reviews Physics}\ }\textbf
  {\bibinfo {volume} {2}},\ \bibinfo {pages} {411} (\bibinfo {year}
  {2020})}\BibitemShut {NoStop}%
\bibitem [{\citenamefont {Adams}\ \emph {et~al.}(2019)\citenamefont {Adams},
  \citenamefont {Pritchard},\ and\ \citenamefont
  {Shaffer}}]{Adams_2020_Rydbergatom}%
  \BibitemOpen
  \bibfield  {author} {\bibinfo {author} {\bibfnamefont {C.~S.}\ \bibnamefont
  {Adams}}, \bibinfo {author} {\bibfnamefont {J.~D.}\ \bibnamefont
  {Pritchard}},\ and\ \bibinfo {author} {\bibfnamefont {J.~P.}\ \bibnamefont
  {Shaffer}},\ }\bibfield  {title} {\bibinfo {title} {{Rydberg atom quantum
  technologies}},\ }\href {https://doi.org/10.1088/1361-6455/ab52ef} {\bibfield
   {journal} {\bibinfo  {journal} {Journal of Physics B: Atomic, Molecular and
  Optical Physics}\ }\textbf {\bibinfo {volume} {53}},\ \bibinfo {pages}
  {012002} (\bibinfo {year} {2019})}\BibitemShut {NoStop}%
\bibitem [{\citenamefont {Browaeys}\ and\ \citenamefont
  {Lahaye}(2020)}]{browaeys2020many}%
  \BibitemOpen
  \bibfield  {author} {\bibinfo {author} {\bibfnamefont {A.}~\bibnamefont
  {Browaeys}}\ and\ \bibinfo {author} {\bibfnamefont {T.}~\bibnamefont
  {Lahaye}},\ }\bibfield  {title} {\bibinfo {title} {{Many-body physics with
  individually controlled Rydberg atoms}},\ }\href
  {https://doi.org/10.1038/s41567-019-0733-z} {\bibfield  {journal} {\bibinfo
  {journal} {Nature Physics}\ }\textbf {\bibinfo {volume} {16}},\ \bibinfo
  {pages} {132} (\bibinfo {year} {2020})}\BibitemShut {NoStop}%
\bibitem [{\citenamefont {Wu}\ \emph {et~al.}(2021)\citenamefont {Wu},
  \citenamefont {Liang}, \citenamefont {Tian}, \citenamefont {Yang},
  \citenamefont {Chen}, \citenamefont {Liu}, \citenamefont {Tey},\ and\
  \citenamefont {You}}]{Wu_2021_Rydbergatom}%
  \BibitemOpen
  \bibfield  {author} {\bibinfo {author} {\bibfnamefont {X.}~\bibnamefont
  {Wu}}, \bibinfo {author} {\bibfnamefont {X.}~\bibnamefont {Liang}}, \bibinfo
  {author} {\bibfnamefont {Y.}~\bibnamefont {Tian}}, \bibinfo {author}
  {\bibfnamefont {F.}~\bibnamefont {Yang}}, \bibinfo {author} {\bibfnamefont
  {C.}~\bibnamefont {Chen}}, \bibinfo {author} {\bibfnamefont {Y.-C.}\
  \bibnamefont {Liu}}, \bibinfo {author} {\bibfnamefont {M.~K.}\ \bibnamefont
  {Tey}},\ and\ \bibinfo {author} {\bibfnamefont {L.}~\bibnamefont {You}},\
  }\bibfield  {title} {\bibinfo {title} {A concise review of {Rydberg} atom
  based quantum computation and quantum simulation*},\ }\href
  {https://doi.org/10.1088/1674-1056/abd76f} {\bibfield  {journal} {\bibinfo
  {journal} {Chinese Physics B}\ }\textbf {\bibinfo {volume} {30}},\ \bibinfo
  {pages} {020305} (\bibinfo {year} {2021})}\BibitemShut {NoStop}%
\bibitem [{\citenamefont {Blatt}\ and\ \citenamefont
  {Roos}(2012)}]{blatt2012quantum}%
  \BibitemOpen
  \bibfield  {author} {\bibinfo {author} {\bibfnamefont {R.}~\bibnamefont
  {Blatt}}\ and\ \bibinfo {author} {\bibfnamefont {C.~F.}\ \bibnamefont
  {Roos}},\ }\bibfield  {title} {\bibinfo {title} {{Quantum simulations with
  trapped ions}},\ }\href {https://doi.org/10.1038/nphys2252} {\bibfield
  {journal} {\bibinfo  {journal} {Nature Physics}\ }\textbf {\bibinfo {volume}
  {8}},\ \bibinfo {pages} {277} (\bibinfo {year} {2012})}\BibitemShut {NoStop}%
\bibitem [{\citenamefont {Monroe}\ \emph {et~al.}(2021)\citenamefont {Monroe},
  \citenamefont {Campbell}, \citenamefont {Duan}, \citenamefont {Gong},
  \citenamefont {Gorshkov}, \citenamefont {Hess}, \citenamefont {Islam},
  \citenamefont {Kim}, \citenamefont {Linke}, \citenamefont {Pagano},
  \citenamefont {Richerme}, \citenamefont {Senko},\ and\ \citenamefont
  {Yao}}]{2021Monroe_trappedions}%
  \BibitemOpen
  \bibfield  {author} {\bibinfo {author} {\bibfnamefont {C.}~\bibnamefont
  {Monroe}}, \bibinfo {author} {\bibfnamefont {W.~C.}\ \bibnamefont
  {Campbell}}, \bibinfo {author} {\bibfnamefont {L.-M.}\ \bibnamefont {Duan}},
  \bibinfo {author} {\bibfnamefont {Z.-X.}\ \bibnamefont {Gong}}, \bibinfo
  {author} {\bibfnamefont {A.~V.}\ \bibnamefont {Gorshkov}}, \bibinfo {author}
  {\bibfnamefont {P.~W.}\ \bibnamefont {Hess}}, \bibinfo {author}
  {\bibfnamefont {R.}~\bibnamefont {Islam}}, \bibinfo {author} {\bibfnamefont
  {K.}~\bibnamefont {Kim}}, \bibinfo {author} {\bibfnamefont {N.~M.}\
  \bibnamefont {Linke}}, \bibinfo {author} {\bibfnamefont {G.}~\bibnamefont
  {Pagano}}, \bibinfo {author} {\bibfnamefont {P.}~\bibnamefont {Richerme}},
  \bibinfo {author} {\bibfnamefont {C.}~\bibnamefont {Senko}},\ and\ \bibinfo
  {author} {\bibfnamefont {N.~Y.}\ \bibnamefont {Yao}},\ }\bibfield  {title}
  {\bibinfo {title} {Programmable quantum simulations of spin systems with
  trapped ions},\ }\href {https://doi.org/10.1103/RevModPhys.93.025001}
  {\bibfield  {journal} {\bibinfo  {journal} {Rev. Mod. Phys.}\ }\textbf
  {\bibinfo {volume} {93}},\ \bibinfo {pages} {025001} (\bibinfo {year}
  {2021})}\BibitemShut {NoStop}%
\bibitem [{\citenamefont {Zeiher}\ \emph {et~al.}(2016)\citenamefont {Zeiher},
  \citenamefont {Van~Bijnen}, \citenamefont {Schau{\ss}}, \citenamefont {Hild},
  \citenamefont {Choi}, \citenamefont {Pohl}, \citenamefont {Bloch},\ and\
  \citenamefont {Gross}}]{zeiher2016many}%
  \BibitemOpen
  \bibfield  {author} {\bibinfo {author} {\bibfnamefont {J.}~\bibnamefont
  {Zeiher}}, \bibinfo {author} {\bibfnamefont {R.}~\bibnamefont {Van~Bijnen}},
  \bibinfo {author} {\bibfnamefont {P.}~\bibnamefont {Schau{\ss}}}, \bibinfo
  {author} {\bibfnamefont {S.}~\bibnamefont {Hild}}, \bibinfo {author}
  {\bibfnamefont {J.-y.}\ \bibnamefont {Choi}}, \bibinfo {author}
  {\bibfnamefont {T.}~\bibnamefont {Pohl}}, \bibinfo {author} {\bibfnamefont
  {I.}~\bibnamefont {Bloch}},\ and\ \bibinfo {author} {\bibfnamefont
  {C.}~\bibnamefont {Gross}},\ }\bibfield  {title} {\bibinfo {title}
  {{Many-body interferometry of a Rydberg-dressed spin lattice}},\ }\href
  {https://doi.org/10.1038/nphys3835} {\bibfield  {journal} {\bibinfo
  {journal} {Nature Physics}\ }\textbf {\bibinfo {volume} {12}},\ \bibinfo
  {pages} {1095} (\bibinfo {year} {2016})}\BibitemShut {NoStop}%
\bibitem [{\citenamefont {Labuhn}\ \emph {et~al.}(2016)\citenamefont {Labuhn},
  \citenamefont {Barredo}, \citenamefont {Ravets}, \citenamefont
  {De~L{\'e}s{\'e}leuc}, \citenamefont {Macr{\`\i}}, \citenamefont {Lahaye},\
  and\ \citenamefont {Browaeys}}]{labuhn2016tunable}%
  \BibitemOpen
  \bibfield  {author} {\bibinfo {author} {\bibfnamefont {H.}~\bibnamefont
  {Labuhn}}, \bibinfo {author} {\bibfnamefont {D.}~\bibnamefont {Barredo}},
  \bibinfo {author} {\bibfnamefont {S.}~\bibnamefont {Ravets}}, \bibinfo
  {author} {\bibfnamefont {S.}~\bibnamefont {De~L{\'e}s{\'e}leuc}}, \bibinfo
  {author} {\bibfnamefont {T.}~\bibnamefont {Macr{\`\i}}}, \bibinfo {author}
  {\bibfnamefont {T.}~\bibnamefont {Lahaye}},\ and\ \bibinfo {author}
  {\bibfnamefont {A.}~\bibnamefont {Browaeys}},\ }\bibfield  {title} {\bibinfo
  {title} {Tunable two-dimensional arrays of single {Rydberg} atoms for
  realizing quantum ising models},\ }\href
  {https://doi.org/10.1038/nature18274} {\bibfield  {journal} {\bibinfo
  {journal} {Nature}\ }\textbf {\bibinfo {volume} {534}},\ \bibinfo {pages}
  {667} (\bibinfo {year} {2016})}\BibitemShut {NoStop}%
\bibitem [{\citenamefont {Zeiher}\ \emph {et~al.}(2017)\citenamefont {Zeiher},
  \citenamefont {Choi}, \citenamefont {Rubio-Abadal}, \citenamefont {Pohl},
  \citenamefont {van Bijnen}, \citenamefont {Bloch},\ and\ \citenamefont
  {Gross}}]{Zeiher2017prx}%
  \BibitemOpen
  \bibfield  {author} {\bibinfo {author} {\bibfnamefont {J.}~\bibnamefont
  {Zeiher}}, \bibinfo {author} {\bibfnamefont {J.-y.}\ \bibnamefont {Choi}},
  \bibinfo {author} {\bibfnamefont {A.}~\bibnamefont {Rubio-Abadal}}, \bibinfo
  {author} {\bibfnamefont {T.}~\bibnamefont {Pohl}}, \bibinfo {author}
  {\bibfnamefont {R.}~\bibnamefont {van Bijnen}}, \bibinfo {author}
  {\bibfnamefont {I.}~\bibnamefont {Bloch}},\ and\ \bibinfo {author}
  {\bibfnamefont {C.}~\bibnamefont {Gross}},\ }\bibfield  {title} {\bibinfo
  {title} {{Coherent Many-Body Spin Dynamics in a Long-Range Interacting Ising
  Chain}},\ }\href {https://doi.org/10.1103/PhysRevX.7.041063} {\bibfield
  {journal} {\bibinfo  {journal} {Phys. Rev. X}\ }\textbf {\bibinfo {volume}
  {7}},\ \bibinfo {pages} {041063} (\bibinfo {year} {2017})}\BibitemShut
  {NoStop}%
\bibitem [{\citenamefont {Scholl}\ \emph {et~al.}(2022)\citenamefont {Scholl},
  \citenamefont {Williams}, \citenamefont {Bornet}, \citenamefont {Wallner},
  \citenamefont {Barredo}, \citenamefont {Henriet}, \citenamefont {Signoles},
  \citenamefont {Hainaut}, \citenamefont {Franz}, \citenamefont {Geier},
  \citenamefont {Tebben}, \citenamefont {Salzinger}, \citenamefont {Z\"urn},
  \citenamefont {Lahaye}, \citenamefont {Weidem\"uller},\ and\ \citenamefont
  {Browaeys}}]{Scholl2022PRXQuantum}%
  \BibitemOpen
  \bibfield  {author} {\bibinfo {author} {\bibfnamefont {P.}~\bibnamefont
  {Scholl}}, \bibinfo {author} {\bibfnamefont {H.~J.}\ \bibnamefont
  {Williams}}, \bibinfo {author} {\bibfnamefont {G.}~\bibnamefont {Bornet}},
  \bibinfo {author} {\bibfnamefont {F.}~\bibnamefont {Wallner}}, \bibinfo
  {author} {\bibfnamefont {D.}~\bibnamefont {Barredo}}, \bibinfo {author}
  {\bibfnamefont {L.}~\bibnamefont {Henriet}}, \bibinfo {author} {\bibfnamefont
  {A.}~\bibnamefont {Signoles}}, \bibinfo {author} {\bibfnamefont
  {C.}~\bibnamefont {Hainaut}}, \bibinfo {author} {\bibfnamefont
  {T.}~\bibnamefont {Franz}}, \bibinfo {author} {\bibfnamefont
  {S.}~\bibnamefont {Geier}}, \bibinfo {author} {\bibfnamefont
  {A.}~\bibnamefont {Tebben}}, \bibinfo {author} {\bibfnamefont
  {A.}~\bibnamefont {Salzinger}}, \bibinfo {author} {\bibfnamefont
  {G.}~\bibnamefont {Z\"urn}}, \bibinfo {author} {\bibfnamefont
  {T.}~\bibnamefont {Lahaye}}, \bibinfo {author} {\bibfnamefont
  {M.}~\bibnamefont {Weidem\"uller}},\ and\ \bibinfo {author} {\bibfnamefont
  {A.}~\bibnamefont {Browaeys}},\ }\bibfield  {title} {\bibinfo {title}
  {{Microwave Engineering of Programmable $XXZ$ Hamiltonians in Arrays of
  Rydberg Atoms}},\ }\href {https://doi.org/10.1103/PRXQuantum.3.020303}
  {\bibfield  {journal} {\bibinfo  {journal} {PRX Quantum}\ }\textbf {\bibinfo
  {volume} {3}},\ \bibinfo {pages} {020303} (\bibinfo {year}
  {2022})}\BibitemShut {NoStop}%
\bibitem [{\citenamefont {Ockeloen}\ \emph {et~al.}(2010)\citenamefont
  {Ockeloen}, \citenamefont {Tauschinsky}, \citenamefont {Spreeuw},\ and\
  \citenamefont {Whitlock}}]{Ockeloen2010_pra}%
  \BibitemOpen
  \bibfield  {author} {\bibinfo {author} {\bibfnamefont {C.~F.}\ \bibnamefont
  {Ockeloen}}, \bibinfo {author} {\bibfnamefont {A.~F.}\ \bibnamefont
  {Tauschinsky}}, \bibinfo {author} {\bibfnamefont {R.~J.~C.}\ \bibnamefont
  {Spreeuw}},\ and\ \bibinfo {author} {\bibfnamefont {S.}~\bibnamefont
  {Whitlock}},\ }\bibfield  {title} {\bibinfo {title} {{Detection of small atom
  numbers through image processing}},\ }\href
  {https://doi.org/10.1103/PhysRevA.82.061606} {\bibfield  {journal} {\bibinfo
  {journal} {Phys. Rev. A}\ }\textbf {\bibinfo {volume} {82}},\ \bibinfo
  {pages} {061606} (\bibinfo {year} {2010})}\BibitemShut {NoStop}%
\bibitem [{\citenamefont {Muessel}\ \emph {et~al.}(2013)\citenamefont
  {Muessel}, \citenamefont {Strobel}, \citenamefont {Joos}, \citenamefont
  {Nicklas}, \citenamefont {Stroescu}, \citenamefont {Tomkovi{\v{c}}},
  \citenamefont {Hume},\ and\ \citenamefont
  {Oberthaler}}]{muessel2013optimized}%
  \BibitemOpen
  \bibfield  {author} {\bibinfo {author} {\bibfnamefont {W.}~\bibnamefont
  {Muessel}}, \bibinfo {author} {\bibfnamefont {H.}~\bibnamefont {Strobel}},
  \bibinfo {author} {\bibfnamefont {M.}~\bibnamefont {Joos}}, \bibinfo {author}
  {\bibfnamefont {E.}~\bibnamefont {Nicklas}}, \bibinfo {author} {\bibfnamefont
  {I.}~\bibnamefont {Stroescu}}, \bibinfo {author} {\bibfnamefont
  {J.}~\bibnamefont {Tomkovi{\v{c}}}}, \bibinfo {author} {\bibfnamefont
  {D.~B.}\ \bibnamefont {Hume}},\ and\ \bibinfo {author} {\bibfnamefont
  {M.~K.}\ \bibnamefont {Oberthaler}},\ }\bibfield  {title} {\bibinfo {title}
  {{Optimized absorption imaging of mesoscopic atomic clouds}},\ }\href
  {https://doi.org/10.1007/s00340-013-5553-8} {\bibfield  {journal} {\bibinfo
  {journal} {Applied Physics B}\ }\textbf {\bibinfo {volume} {113}},\ \bibinfo
  {pages} {69} (\bibinfo {year} {2013})}\BibitemShut {NoStop}%
\bibitem [{\citenamefont {Hume}\ \emph {et~al.}(2013)\citenamefont {Hume},
  \citenamefont {Stroescu}, \citenamefont {Joos}, \citenamefont {Muessel},
  \citenamefont {Strobel},\ and\ \citenamefont {Oberthaler}}]{Hume2013prl}%
  \BibitemOpen
  \bibfield  {author} {\bibinfo {author} {\bibfnamefont {D.~B.}\ \bibnamefont
  {Hume}}, \bibinfo {author} {\bibfnamefont {I.}~\bibnamefont {Stroescu}},
  \bibinfo {author} {\bibfnamefont {M.}~\bibnamefont {Joos}}, \bibinfo {author}
  {\bibfnamefont {W.}~\bibnamefont {Muessel}}, \bibinfo {author} {\bibfnamefont
  {H.}~\bibnamefont {Strobel}},\ and\ \bibinfo {author} {\bibfnamefont {M.~K.}\
  \bibnamefont {Oberthaler}},\ }\bibfield  {title} {\bibinfo {title} {{Accurate
  Atom Counting in Mesoscopic Ensembles}},\ }\href
  {https://doi.org/10.1103/PhysRevLett.111.253001} {\bibfield  {journal}
  {\bibinfo  {journal} {Phys. Rev. Lett.}\ }\textbf {\bibinfo {volume} {111}},\
  \bibinfo {pages} {253001} (\bibinfo {year} {2013})}\BibitemShut {NoStop}%
\bibitem [{\citenamefont {Dimitrova}\ \emph {et~al.}(2023)\citenamefont
  {Dimitrova}, \citenamefont {Flannigan}, \citenamefont {Lee}, \citenamefont
  {Lin}, \citenamefont {Amato-Grill}, \citenamefont {Jepsen}, \citenamefont
  {Čepaitė}, \citenamefont {Daley},\ and\ \citenamefont
  {Ketterle}}]{Dimitrova_2023}%
  \BibitemOpen
  \bibfield  {author} {\bibinfo {author} {\bibfnamefont {I.}~\bibnamefont
  {Dimitrova}}, \bibinfo {author} {\bibfnamefont {S.}~\bibnamefont
  {Flannigan}}, \bibinfo {author} {\bibfnamefont {Y.~K.}\ \bibnamefont {Lee}},
  \bibinfo {author} {\bibfnamefont {H.}~\bibnamefont {Lin}}, \bibinfo {author}
  {\bibfnamefont {J.}~\bibnamefont {Amato-Grill}}, \bibinfo {author}
  {\bibfnamefont {N.}~\bibnamefont {Jepsen}}, \bibinfo {author} {\bibfnamefont
  {I.}~\bibnamefont {Čepaitė}}, \bibinfo {author} {\bibfnamefont {A.~J.}\
  \bibnamefont {Daley}},\ and\ \bibinfo {author} {\bibfnamefont
  {W.}~\bibnamefont {Ketterle}},\ }\bibfield  {title} {\bibinfo {title}
  {{Many-body spin rotation by adiabatic passage in spin-1/2 XXZ chains of
  ultracold atoms}},\ }\href {https://doi.org/10.1088/2058-9565/acd2fb}
  {\bibfield  {journal} {\bibinfo  {journal} {Quantum Science and Technology}\
  }\textbf {\bibinfo {volume} {8}},\ \bibinfo {pages} {035018} (\bibinfo {year}
  {2023})}\BibitemShut {NoStop}%
\bibitem [{\citenamefont {Colombo}\ \emph {et~al.}(2022)\citenamefont
  {Colombo}, \citenamefont {Pedrozo-Penafiel}, \citenamefont {Adiyatullin},
  \citenamefont {Li}, \citenamefont {Mendez}, \citenamefont {Shu},\ and\
  \citenamefont {Vuleti{\'c}}}]{colombo2022time}%
  \BibitemOpen
  \bibfield  {author} {\bibinfo {author} {\bibfnamefont {S.}~\bibnamefont
  {Colombo}}, \bibinfo {author} {\bibfnamefont {E.}~\bibnamefont
  {Pedrozo-Penafiel}}, \bibinfo {author} {\bibfnamefont {A.~F.}\ \bibnamefont
  {Adiyatullin}}, \bibinfo {author} {\bibfnamefont {Z.}~\bibnamefont {Li}},
  \bibinfo {author} {\bibfnamefont {E.}~\bibnamefont {Mendez}}, \bibinfo
  {author} {\bibfnamefont {C.}~\bibnamefont {Shu}},\ and\ \bibinfo {author}
  {\bibfnamefont {V.}~\bibnamefont {Vuleti{\'c}}},\ }\bibfield  {title}
  {\bibinfo {title} {{Time-reversal-based quantum metrology with many-body
  entangled states}},\ }\href {https://doi.org/10.1038/s41567-022-01653-5}
  {\bibfield  {journal} {\bibinfo  {journal} {Nature Physics}\ }\textbf
  {\bibinfo {volume} {18}},\ \bibinfo {pages} {925} (\bibinfo {year}
  {2022})}\BibitemShut {NoStop}%
\bibitem [{\citenamefont {Bornet}\ \emph {et~al.}(2023)\citenamefont {Bornet},
  \citenamefont {Emperauger}, \citenamefont {Chen}, \citenamefont {Ye},
  \citenamefont {Block}, \citenamefont {Bintz}, \citenamefont {Boyd},
  \citenamefont {Barredo}, \citenamefont {Comparin}, \citenamefont {Mezzacapo}
  \emph {et~al.}}]{bornet2023scalable}%
  \BibitemOpen
  \bibfield  {author} {\bibinfo {author} {\bibfnamefont {G.}~\bibnamefont
  {Bornet}}, \bibinfo {author} {\bibfnamefont {G.}~\bibnamefont {Emperauger}},
  \bibinfo {author} {\bibfnamefont {C.}~\bibnamefont {Chen}}, \bibinfo {author}
  {\bibfnamefont {B.}~\bibnamefont {Ye}}, \bibinfo {author} {\bibfnamefont
  {M.}~\bibnamefont {Block}}, \bibinfo {author} {\bibfnamefont
  {M.}~\bibnamefont {Bintz}}, \bibinfo {author} {\bibfnamefont {J.~A.}\
  \bibnamefont {Boyd}}, \bibinfo {author} {\bibfnamefont {D.}~\bibnamefont
  {Barredo}}, \bibinfo {author} {\bibfnamefont {T.}~\bibnamefont {Comparin}},
  \bibinfo {author} {\bibfnamefont {F.}~\bibnamefont {Mezzacapo}}, \emph
  {et~al.},\ }\bibfield  {title} {\bibinfo {title} {{Scalable spin squeezing in
  a dipolar Rydberg atom array}},\ }\href
  {https://doi.org/10.1038/s41586-023-06414-9} {\bibfield  {journal} {\bibinfo
  {journal} {Nature}\ }\textbf {\bibinfo {volume} {621}},\ \bibinfo {pages}
  {728} (\bibinfo {year} {2023})}\BibitemShut {NoStop}%
\bibitem [{\citenamefont {Franke}\ \emph {et~al.}(2023)\citenamefont {Franke},
  \citenamefont {Muleady}, \citenamefont {Kaubruegger}, \citenamefont {Kranzl},
  \citenamefont {Blatt}, \citenamefont {Rey}, \citenamefont {Joshi},\ and\
  \citenamefont {Roos}}]{franke2023quantum}%
  \BibitemOpen
  \bibfield  {author} {\bibinfo {author} {\bibfnamefont {J.}~\bibnamefont
  {Franke}}, \bibinfo {author} {\bibfnamefont {S.~R.}\ \bibnamefont {Muleady}},
  \bibinfo {author} {\bibfnamefont {R.}~\bibnamefont {Kaubruegger}}, \bibinfo
  {author} {\bibfnamefont {F.}~\bibnamefont {Kranzl}}, \bibinfo {author}
  {\bibfnamefont {R.}~\bibnamefont {Blatt}}, \bibinfo {author} {\bibfnamefont
  {A.~M.}\ \bibnamefont {Rey}}, \bibinfo {author} {\bibfnamefont {M.~K.}\
  \bibnamefont {Joshi}},\ and\ \bibinfo {author} {\bibfnamefont {C.~F.}\
  \bibnamefont {Roos}},\ }\bibfield  {title} {\bibinfo {title}
  {{Quantum-enhanced sensing on optical transitions through finite-range
  interactions}},\ }\href {https://doi.org/10.1038/s41586-023-06472-z}
  {\bibfield  {journal} {\bibinfo  {journal} {Nature}\ }\textbf {\bibinfo
  {volume} {621}},\ \bibinfo {pages} {740} (\bibinfo {year}
  {2023})}\BibitemShut {NoStop}%
\bibitem [{\citenamefont {Imai}\ \emph {et~al.}(2024)\citenamefont {Imai},
  \citenamefont {T\'oth},\ and\ \citenamefont {G\"uhne}}]{Imai2024prl}%
  \BibitemOpen
  \bibfield  {author} {\bibinfo {author} {\bibfnamefont {S.}~\bibnamefont
  {Imai}}, \bibinfo {author} {\bibfnamefont {G.}~\bibnamefont {T\'oth}},\ and\
  \bibinfo {author} {\bibfnamefont {O.}~\bibnamefont {G\"uhne}},\ }\bibfield
  {title} {\bibinfo {title} {{Collective Randomized Measurements in Quantum
  Information Processing}},\ }\href
  {https://doi.org/10.1103/PhysRevLett.133.060203} {\bibfield  {journal}
  {\bibinfo  {journal} {Phys. Rev. Lett.}\ }\textbf {\bibinfo {volume} {133}},\
  \bibinfo {pages} {060203} (\bibinfo {year} {2024})}\BibitemShut {NoStop}%
\end{thebibliography}
\end{document}